\documentclass{SciPost}

\hypersetup{
    colorlinks,
    linkcolor={red!50!black},
    citecolor={blue!50!black},
    urlcolor={blue!80!black}
}

\usepackage[bitstream-charter]{mathdesign}
\DeclareSymbolFont{usualmathcal}{OMS}{cmsy}{m}{n}
\DeclareSymbolFontAlphabet{\mathcal}{usualmathcal}

\fancypagestyle{SPstyle}{
\fancyhf{}
\lhead{\colorbox{scipostblue}{\bf \color{white} ~SciPost Physics }}
\rhead{{\bf \color{scipostdeepblue} ~Submission }}

\fancyfoot[C]{\textbf{\thepage}}
}

\usepackage{lipsum}
\usepackage{braket}
\usepackage{graphicx}
\usepackage{bm}
\usepackage{tabularx}
\usepackage{amsmath}

\usepackage[normalem]{ulem}

\let\linenumbers\relax

\DeclareMathOperator{\Tr}{Tr}

\newcommand{\RN}[1]{%
	\textup{\uppercase\expandafter{\romannumeral#1}}%
}
\newcommand{\Rn}[1]{%
	\textup{\lowercase\expandafter{\romannumeral#1}}%
}

\begin{document}

\pagestyle{SPstyle}

\begin{center}{\Large \textbf{\color{scipostdeepblue}{
Generalized Loschmidt echo and information scrambling in open systems\\
}}}\end{center}

\begin{center}\textbf{
Yi-Neng Zhou\textsuperscript{1,2$\star$} and
Chang Liu\textsuperscript{1,3$\dagger$}
}\end{center}

\begin{center}
{\bf 1} Institute for Advanced Study, Tsinghua University, Beijing,100084, China
\\
{\bf 2} Department of Theoretical Physics, University of Genève, 1211 Genève 4, Suisse
\\
{\bf 3} Department of Physics, National University of Singapore, Singapore 117542
\\[\baselineskip]
$\star$ \href{zhouyn.physics@gmail.com}{\small zhouyn.physics@gmail.com}\,,\quad
$\dagger$ \href{changliu@nus.edu.sg}{\small changliu@nus.edu.sg}
\end{center}

\section*{\color{scipostdeepblue}{Abstract}}
\textbf{\boldmath{%
Quantum information scrambling, typically explored in closed quantum systems, describes the spread of initially localized information throughout a system and can be quantified by measures such as the Loschmidt echo (LE) and out-of-time-order correlator (OTOC). In this paper, we explore information scrambling in the presence of dissipation by generalizing the concepts of LE and OTOC to open quantum systems governed by Lindblad dynamics.  We investigate the universal dynamics of the generalized LE across regimes of weak and strong dissipation. In the weak dissipation regime, we identify a universal structure, while in the strong dissipation regime, we observe a distinctive two-local-minima structure, which we interpret through an analysis of the Lindblad spectrum. Furthermore, we establish connections between the thermal averages of LE and OTOC and prove a general relation between OTOC and R\'enyi entropy in open systems. Finally, we propose an experimental protocol for measuring OTOC in open systems. These findings provide deeper insights into information scrambling under dissipation and pave the way for experimental studies in open quantum systems.
}}

\vspace{\baselineskip}

\noindent\textcolor{white!90!black}{%
\fbox{\parbox{0.975\linewidth}{%
\textcolor{white!40!black}{\begin{tabular}{lr}%
  \begin{minipage}{0.6\textwidth}%
    {\small Copyright attribution to authors. \newline
    This work is a submission to SciPost Physics. \newline
    License information to appear upon publication. \newline
    Publication information to appear upon publication.}
  \end{minipage} & \begin{minipage}{0.4\textwidth}
    {\small Received Date \newline Accepted Date \newline Published Date}%
  \end{minipage}
\end{tabular}}
}}
}

\linenumbers

\vspace{10pt}
\noindent\rule{\textwidth}{1pt}
\tableofcontents
\noindent\rule{\textwidth}{1pt}
\vspace{10pt}


\section{Introduction}
\label{sec:intro}
Quantum information scrambling has become a pivotal concept in understanding the dynamics of quantum many-body systems, attracting significant attention in recent years. In closed quantum systems, it describes the process by which initially localized quantum information spreads throughout the system, rendering it inaccessible to local measurements. This phenomenon is closely tied to the thermalization dynamics of many-body systems and serves as a key diagnostic for quantum chaos.

One of the most prominent tools for probing quantum information scrambling is the out-of-time-order correlator (OTOC)~\cite{larkin1969quasiclassical,kitaev2015simple}, which has been extensively studied across various areas of physics~\cite{almheiriApologiaFirewalls2013,shenkerBlackHolesButterfly2014,robertsDiagnosingChaosUsing2015,robertsLocalizedShocks2015,shenkerStringyEffectsScrambling2015,hosurChaosQuantumChannels2016a,Maldacena_2016,stanfordManybodyChaosWeak2016a,maldacenaConformalSymmetryIts2016,maldacenaRemarksSachdevYeKitaevModel2016a,zhuMeasurementManybodyChaos2016,StanfordLocalCriticalityDiffusion2017,ReyMeasuringOutoftimeorderCorrelations2017,chenTunableQuantumChaos2017a,ZhangCompetitionChaoticNonchaotic2017,BalentsStronglyCorrelatedMetal2017,patelQuantumButterflyEffect2017,vonkeyserlingkOperatorHydrodynamicsOTOCs2018a,xuLocalityQuantumFluctuations2019a,zhangQuantumChaosUnitary2019a,KitaevRelationMagnitudeExponent2019,KitaevObstacleSubAdSHolography2021,SachdevTransportChaosLattice2019,SuhSoftModeSachdevYeKitaev2018,KitaevObstacleSubAdSHolography2021,ZhangTwowayApproachOutoftimeorder2022,Tian_2022,Zamani_2022} and has recently been measured in experiments~\cite{PhysRevX.7.031011,PhysRevLett.120.070501,PhysRevLett.125.120504, Braum_ller_2021, Mi_2021,G_rttner_2017,Landsman_2019,Joshi_2020, Pegahan_2021}. Another quantity closely related to the dynamics of information scrambling is the Loschmidt echo (LE)~\cite{PhysRevA.30.1610,PASTAWSKI2000166,PhysRevLett.86.2490,gorin2006dynamics,doi:10.1080/00018730902831009,goussev2012loschmidt}, which characterizes the retrieval fidelity of a quantum state after an imperfect time-reversal procedure, and has also been measured in experiments~\cite{PhysRev.80.580,PhysRevLett.25.218,PhysRevB.86.214410,PhysRevLett.69.2149,levsteinAttenuationPolarizationEchoes1998,PhysRevLett.124.030601,PhysRevA.105.052232,PhysRevA.104.012402,sanchez2020perturbation,sanchez2022emergent,dominguez2021decoherence}. The LE reveals that even small perturbations in the Hamiltonian can lead to significant changes in the system's dynamics, a hallmark of quantum chaotic behavior.

In closed quantum systems, information scrambling has been extensively studied, with general relations between quantities such as the LE and the OTOC being widely explored~\cite{Maldacena_2016,fanOutofTimeOrderCorrelationManyBody2017,yanInformationScramblingLoschmidt2020, Kudler_Flam_2020,zhang2019information,yoshida2019disentangling,dominguez2021decoherence,xu2021thermofield,xu2019extreme,tuziemski2019out,syzranov2018out,zanardi2021information,swingle2018resilience,PhysRevA.103.062214,zhou2025measuringrenyientropyusing}. However, in realistic experimental settings, interactions between the system and its environment are unavoidable. The presence of dissipation and decoherence significantly modifies the universal behavior of information scrambling, making it fundamentally different in open quantum systems~\cite{Landsman_2019,Mi_2021,arute2019quantum,gonzalez2019out,yoshida2019disentangling,swingle2018resilience,vermersch2019probing,agarwal2020toy,touil2021information,andreadakis2023scrambling,harris2022benchmarking}. Also, generalizing familiar concepts from closed systems to open systems has led to the discovery of novel and intriguing physics including the development of entropy dynamics \cite{zhouEnyiEntropyDynamics2021,zhou2023generalizeda,PhysRevD.109.L081901,kobayashi2024timeevolutionvonneumann}, operator growth and complexity \cite{bhattacharyaOperatorGrowthKrylov2022, PhysRevLett.131.160402, PhysRevResearch.5.033085, PathakOperatorGrowthOpen2023,bhattacharyaKrylovComplexityOpen2023}, strong to weak symmetry breaking \cite{lee2301quantum,ogunnaike2023unifying,ma2024topologicalphasesaveragesymmetries,lessa2024strongtoweakspontaneoussymmetrybreaking,sala2024spontaneousstrongsymmetrybreaking,xu2024averageexactmixedanomaliescompatible,huang2024hydrodynamicseffectivefieldtheory,Kuno_2024,gu2024spontaneous}, quantum speed limit \cite{PhysRevLett.110.050402, PhysRevLett.110.050403,Deffner_2013,PhysRevX.12.011038,Nakajima_2022,sekiguchi2024improvementspeedlimitsquantum},  generalize Lieb–Schultz–Mattis theorem in open systems \cite{zhou2024reviving,Kawabata_2024}. This raises a natural question: how do the dynamics of OTOC and LE behave in open quantum systems?

In this paper, we generalize the concepts of the LE and the OTOC to open quantum systems governed by the Lindblad master equation. Interaction with the environment enriches the dynamics of information scrambling, creating a complex interplay between quantum chaos and dissipation. We investigate the universal behavior of the LE, analyzing its dynamics in both weak and strong dissipation regimes. In the weak dissipation regime, we identify a universal structure for the LE and explain its characteristic behavior. In the strong dissipation regime, we discover that the LE can exhibit a two-local-minima structure, reflecting the Lindblad spectrum of the dissipative evolution. Additionally, we establish a connection between the thermal average of the LE and the OTOC averaging over all unitary operators on two subsystems; For a large class of density matrices, we also prove a general relation between the R\'enyi entropy  and averaged OTOC for corresponding operators in open systems. Finally, we propose an experimental protocol for measuring the OTOC in open systems.

    The structure of this paper is organized as follows. In Section \ref{Loschmidt_in_open}, we review the definition of the LE in closed systems, introduce its generalization to open systems, and discuss its physical significance. In Section \ref{Loschmidt_dynamics}, we analyze the simplest case where the forward and backward evolutions of the LE in open systems differ only by the dissipation strength, exploring LE dynamics in both weak and strong dissipation regimes. In Section \ref{OTOC-LE_section}, we review the relation between the OTOC and the LE in closed systems, introduce the definition of the OTOC in open systems, and establish a similar OTOC-LE relation for open systems. In Section \ref{OTOC-RE_section}, we extend the relation between the OTOC and R\'enyi entropy from closed systems to open systems. In Section \ref{OTOC_experiment}, we propose an experimental protocol for measuring the OTOC in open quantum systems. Finally, in Section \ref{summary}, we present our conclusions and discuss the implications of our findings.
 
\section{The Loschmidt echo in open systems}\label{Loschmidt_in_open}

\begin{figure}[t] 
	\centering 
	\includegraphics[width=0.9\textwidth]{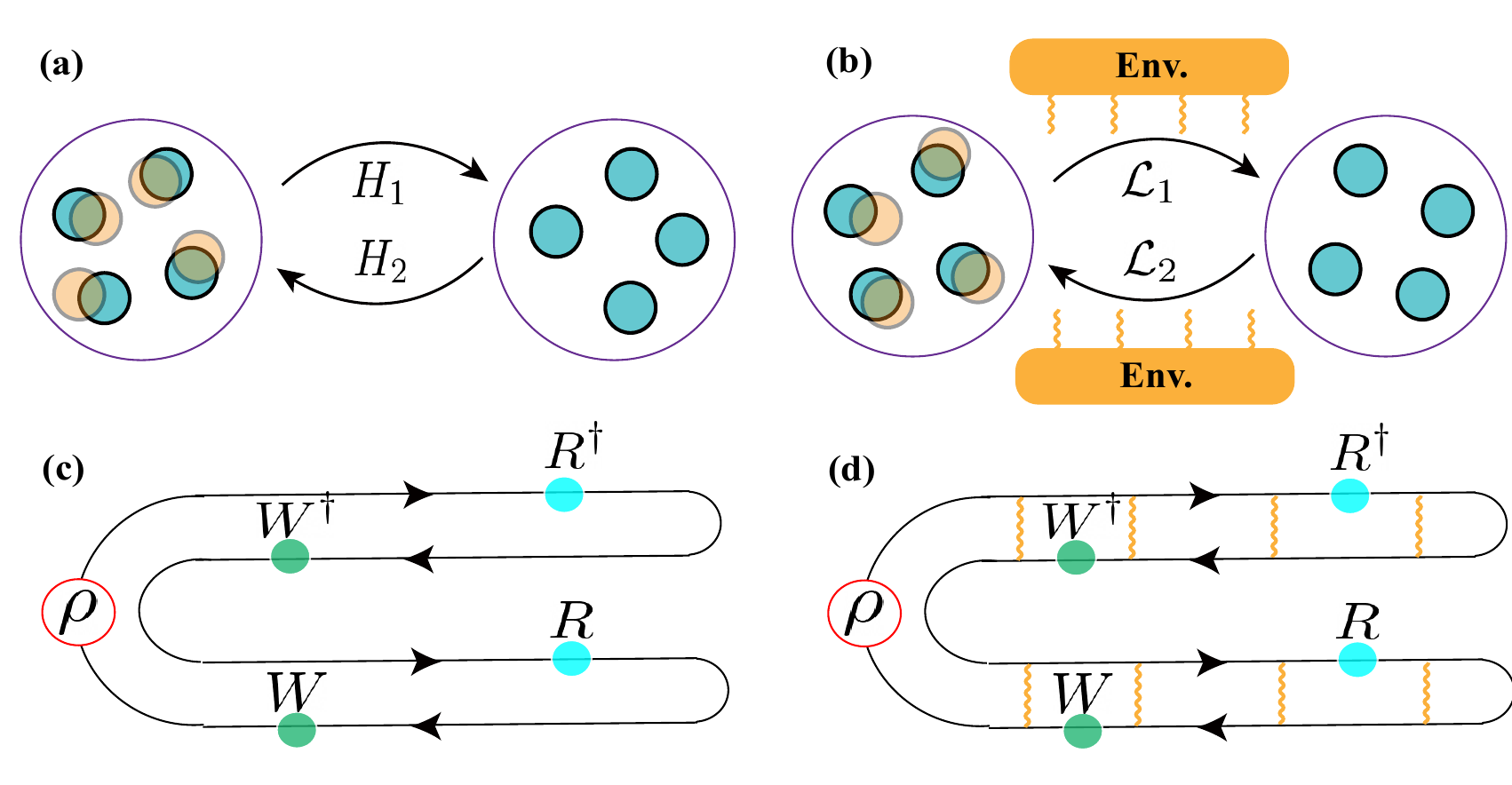} 
	\caption{Schematic of the LE in closed (a) and open systems (b), as well as the  OTOC in closed (c) and open systems (d). (a): In closed systems, the forward and backward time evolutions are governed by different Hamiltonians ($H_1$ and $H_2$), resulting in a final state (orange circle) that differs from the initial state (blue circle). (b): In open systems, due to interactions with the environment (Env.), the Lindbladians $\mathcal{L}_1$ and $\mathcal{L}_2$ govern the forward and backward time evolutions, respectively. (c-d): The time contours of the OTOC. In contrast to the time evolution of OTOC in closed systems (c), the forward and backward time evolution in open systems have correlations, as denoted by the orange curved lines in (d).
 }
	  \label{LE_schematic}
\end{figure}

In an isolated quantum system, the LE measures the retrieval fidelity of a quantum state following an imperfect time-reversal evolution. It is defined as follows:
	\begin{equation}
		M(t)=|\langle \psi_0|e^{iH_2 t} e^{-iH_1t}|\psi_0 \rangle |^2.
	\label{Echo_closed_def}
	\end{equation} 
	Here, $H_1$ and $H_2$ are the Hamiltonians governing the forward and backward time evolution, respectively. $\ket{\psi_0}$ is the initial quantum state at time $t_0 =0$. Consider the case where $H_2=H_1+V$, with $V$ representing a perturbation to $H_1$, thus, the LE measures the sensitivity of quantum evolution to the perturbation and quantifies the degree of irreversibility. Even small perturbations to the Hamiltonian can induce significant changes in the system's dynamics, resulting in a substantial deviation of the LE from 1 as the total evolution time increases from 0. The definition of the LE is illustrated schematically in Fig.~\ref{LE_schematic}(a). In real experiments, the measurement of the LE has been performed in numerous nuclear magnetic resonance (NMR) studies since the 1950s \cite{PhysRev.80.580,PhysRevLett.25.218,PhysRevB.86.214410,PhysRevLett.69.2149,levsteinAttenuationPolarizationEchoes1998,PhysRevLett.124.030601,PhysRevA.105.052232,PhysRevA.104.012402,sanchez2020perturbation,sanchez2022emergent,dominguez2021decoherence}, where echo experiments have served as a standard tool.

We generalize the definition of the LE for open systems and will further discuss its dynamics \cite{gorin2006dynamics,goussev2012loschmidt} in the next section. Specifically, we consider the LE in open quantum systems coupled to a Markovian environment, where the dynamics are governed by the Lindblad master equation.
	\begin{equation} 
		\frac{\partial{\rho}}{\partial t}=-i[H,\rho] +2\gamma\sum_m L_m \rho L^{\dagger}_m-\gamma\sum_m\lbrace L^{\dagger}_m L_m, \rho\rbrace.
	\end{equation}
Here, the first term on the right-hand side (R.H.S.) describes the unitary time evolution with Hamiltonian $H$, while the remaining terms describe the interaction between the system and the environment, characterized by the Lindblad jump operators ${L_m}$ and the dissipation strength $\gamma$. Below, we use $\rho(t) = e^{\mathcal{L}t}[\rho_0]$ to denote the density matrix obtained by evolving the initial density matrix $\rho_0$ under the Lindblad master equation for a total time $t$.

Considering a general density matrix $\rho$, and given a set of complete bases $\{|n\rangle\}$,  the
density matrix $\rho$ can be expressed as 
\begin{equation} 
\rho = \sum_{m,n} \rho_{mn} |m\rangle \langle n|.
\end{equation}
By using the Choi-Jamiolkwski isomorphism \cite{Tyson_2003,Zwolak_2004,Prosen_2008},  the density matrix $\rho$ can be mapped to a  wave function $| \psi_{\rho}^D\rangle $ in double space, where 
\begin{equation} 
| \psi_{\rho}^D\rangle=\sum_{mn}\rho_{mn}|m\rangle_L \otimes |n\rangle_R.
\end{equation}
Here, we denote the two copies of the system in double space as the left (L) and right (R) systems.

After this mapping, the Lindblad master equation, which describes the dynamics of the density matrix, can be transformed into a non-unitary Schrödinger-like equation that governs the dynamics of the double-space wave function $|\psi_{\rho}^D\rangle$ \cite{Tyson_2003,Zwolak_2004,Prosen_2008,zhouEnyiEntropyDynamics2021}: 
 \begin{equation} i\partial_t{|\psi_{\rho}^D(t)\rangle}=H^D|\psi_{\rho}^D(t)\rangle.
	\end{equation}
	Here, $H^D=H_s-iH_d$ is a non-Hermitian operator defined on the double space with 
	\begin{equation} 
		\begin{split}
            H_s =& H_L\otimes \mathcal{I}_R - \mathcal{I}_L\otimes H^T_R,\\
			H_d =&\gamma \sum_m [-2L_{m,L}\otimes L_{m,R}^* +(L^\dagger_mL_m)_L\otimes \mathcal{I}_R +\mathcal{I}_L\otimes (L^\dagger_mL_m)_{R}^*],
		\end{split}
		\label{H_double}
	\end{equation}
	where the superscript $``\text{T}"$ stands for the transpose, and $*$ stands for the complex conjugation. Here, $H_s$ is the Hermitian part of the double space Hamiltonian, determined by the system Hamiltonian. $H_d$ originates from the dissipative part of the Lindblad evolution and is determined by the dissipation operators; hence, we refer to it as the ``dissipative part of the double space Hamiltonian.''

 Using this double-space representation, we can define the generalized LE in open systems as:
	\begin{equation}
		\begin{split}
			&M^D(t)
			\equiv\frac{\Tr \left[\rho_1(t)\rho_2(t)\right] }{\sqrt{\Tr\left[\rho_1^2(t)\right] \Tr\left[\rho_2^2(t)\right]}}=\frac{\langle \psi^D_0 |e^{iH_2^{D \dagger} t} e^{-iH_1^Dt}|\psi^D_0 \rangle }{\sqrt{\langle \psi^D_0|e^{iH_1^{D \dagger} t} e^{-iH_1^D t}|\psi^D_0 \rangle \langle \psi^D_0|e^{iH_2^{D \dagger} t} e^{-iH_2^Dt}|\psi^D_0 \rangle}}.
			\label{Loschmidt_open_def}
		\end{split}
	\end{equation} 
	Here, $\rho_1(t)=e^{\mathcal{L}_1t}[\rho_0]$ and $\rho_2(t)=e^{\mathcal{L}_2t}[\rho_0]$. $\rho_1$ and $\rho_2$ represent the density matrix under the forward Lindblad  superoperator $\mathcal{L}_1$ and backward Lindblad  superoperator $\mathcal{L}_2$ respectively. $|\psi^D_0 \rangle$ is the double space wave function obtained from the initial density matrix $\rho_0$. $H_1^D$ and $H_2^D$ are the double-space Hamiltonians corresponding to the Lindblad  superoperator $\mathcal{L}_1$ and $\mathcal{L}_2$, respectively, as defined in Eq.~\eqref{H_double}. Thus, $H_1^D$ and $H_2^D$ can also be viewed as the forward and backward double-space Hamiltonians.  The illustration of the generalized LE in the open system is depicted in  Fig.~\ref{LE_schematic}(b). 
    This generalized LE is simply the normalized Hilbert--Schmidt overlap between the two time-evolved density matrices $\rho_1(t)$ and $\rho_2(t)$. It quantifies the sensitivity of the open-system dynamics to perturbations of the Lindbladian, i.e., to changes in both the Hamiltonian and the dissipative part. This directly parallels the closed-system LE, which measures the sensitivity of purely unitary dynamics to Hamiltonian perturbations.
 
 Our definition of the generalized LE describes the system's departure from its initial state after undergoing forward lindblad evolution  $\mathcal{L}_1$ followed by a distinct backward lindblad evolution $\mathcal{L}_2$. Since the purity of the density matrix, $\Tr \left[\rho^2(t)\right]$, generally decays due to the decoherence in open systems, even when the forward and backward time evolutions are identical ($\rho_1 = \rho_2$), the overlap $\Tr \left[\rho_1(t)\rho_2(t)\right]$ can still decrease over time. To characterize the intrinsic differences between the forward and backward time evolution in open systems, we introduce the normalization factor $\sqrt{\Tr\left[\rho_1^2(t)\right] \Tr\left[\rho_2^2(t)\right]}$ in Eq.~\eqref{Loschmidt_open_def} to exclude the purity decay of the density matrix during Lindblad time evolution.

  If the dissipation strengths $\gamma_1$ and $\gamma_2$ in the Lindblad  superoperators $\mathcal{L}_1$ and $\mathcal{L}_2$ are set to zero, and the initial double-space wave function $\ket{\psi^D} = |\psi_0\rangle \otimes |\psi_0\rangle$ is purified from a pure state density matrix $\rho_0 = \ket{\psi_0} \bra{\psi_0}$, then the generalized LE defined in Eq.~\eqref{Loschmidt_open_def} can be rewritten as 
		\begin{equation}
		\begin{split}
            M^D(t)|_{\gamma_1=\gamma_2=0}=&\Tr(e^{-iH_1t}\rho_0e^{iH_1t} e^{-iH_2t} \rho_0e^{iH_2t})=|\langle \psi_0|e^{iH_2 t} e^{-iH_1t}|\psi_0 \rangle |^2.
		\end{split}
	\end{equation} 
Hence, in the absence of dissipation, the generalized LE in an open system reduces to its definition in closed systems, as given by Eq.~\eqref{Echo_closed_def}. In this context, it quantifies the quantum fidelity between two states after undergoing different Hamiltonian evolutions.

Additionally, the generalized LE can be interpreted as describing the overlap between two (mixed) density matrices. This overlap quantifies the similarity between the two mixed states and is related to the relative purity, which is defined as
  	\begin{equation}
		F_{RP}(\rho_1,\rho_2)=\frac{\Tr(\rho_1\rho_2)}{\Tr(\rho_1^2)}.
	\end{equation} 
The rate of change of relative purity can signal quantum fluctuations and is often discussed in the context of quantum speed limits \cite{mandelstam1991uncertainty, Margolus_1998, PhysRevLett.103.160502, frey2016quantum, Deffner_2017}. It is worth noting that the only distinction between relative purity and the generalized LE we have defined lies in their respective normalization factors. Therefore, the generalized LE in open systems can also be used to analyze quantum speed limits.

\section{The Loschmidt echo dynamics in open systems } \label{Loschmidt_dynamics}

In open systems, our generalized LE can be used to characterize the irreversibility between different forward and backward dissipative dynamics. In principle, the forward and backward Lindblad evolutions can differ in several aspects: the system Hamiltonian $H$, the dissipation strength $\gamma$, or the set of Lindblad jump operators $\{L_m\}$. In this work, to elucidate the role of dissipation, we consider the simplest case where the forward and backward evolutions share the same Hamiltonians and dissipation operators, differing only in the dissipation strengths $\gamma$.

We investigate the universal dynamical properties of the generalized LE and identify the characteristic time scales that arise from the interplay between the two different dissipation strengths of the Lindblad superoperator and the energy scales of the Hamiltonian.

Without loss of generality, we numerically study a dissipative Sachdev-Ye-Kitaev (SYK) model as an example to illustrate the typical dynamical behavior of the generalized LE in open systems. We consider a quantum system composed of $N$ Majorana fermions, described by the SYK model Hamiltonian~\cite{Sachdev_1993,Kitaev:2015,Sachdev:2015efa,Maldacena:2016hyu},
\begin{equation}
    H_{\text{SYK}} = \sum_{ 1\leq i<j<k<l \leq N}J_{ijkl} \chi_i \chi_j \chi_k \chi_l,
\end{equation}
where  $\chi_i$ denotes the Majorana fermion operator that satisfies the anticommutation relation $\{\chi_i,\chi_j\} = \delta_{ij}$. Here, $J_{ijkl}$ is a random variable that satisfies the Gaussian
distribution with zero mean and variance 
\begin{equation*} 
	\langle J_{ijkl} J_{i^{'}j^{'}k^{'}l^{'}} \rangle=\delta_{i,i^{'}}\delta_{j,j^{'}}\delta_{k,k^{'}}\delta_{l,l^{'}}\frac{3!J^2(q-1)!}{N^{3}}.
\end{equation*}
We consider the set of jump operators $\{L_j = \chi_j\}$, $(j=1,2,...,N)$ with dissipation strength $\gamma$~\cite{kulkarni2022lindbladian},  leading to the Lindblad time evolution:
\begin{equation} 
		\frac{\partial{\rho}}{\partial t}=-i[H_{\text{SYK}},\rho] +2\gamma\sum_{j=1}^N \chi_j \rho \chi_j-\gamma\sum_{j=1}^N\lbrace \chi_j \chi_j, \rho\rbrace.
        \label{dissSYK_all_chain}
	\end{equation}

Below, we use the dissipative SYK model as an example to study the dynamics of the LE in both the weak dissipation regime, where $\gamma/J \ll 1$, and the strong dissipation regime, where $\gamma/J \gg 1$. Our findings are summarized in Table~\ref{tab:addlabel}. Without loss of generality, we denote the dissipation strengths of the forward and backward Lindblad evolutions as $\gamma_1$ and $\gamma_2$, with $\gamma_1 < \gamma_2$. For clarity, we will refer to the Lindblad evolution with dissipation strengths $\gamma_1$ and $\gamma_2$ as the $\gamma_1$ evolution and $\gamma_2$ evolution, respectively.

\begin{table}[h!]
  \centering
    \begin{tabular}{|c|c|c|c|}
    \hline
    Cases & Typical dynamical structure & Late time plateau \\
    \hline
    weak: \\$\gamma_1/J < \gamma_2/J \ll 1$ & One minimum, $t_{\text{min}}\sim\frac{1}{\gamma_2}$ & $t_{\text{p}} \sim 1/\gamma_1$ \\
    \hline
    strong \ (non-deg): \\$1 \ll \gamma_1/J < \gamma_2/J $  & One miniumum, $t_{\text{min}}\sim\frac{1}{\gamma_2}$ & $t_{\text{p}} \sim 1/\gamma_1$ \\
    \hline
     strong \ (degenerate): \\$1 \ll \gamma_1/J < \gamma_2/J $   & Two local minima, $t_{\text{min}1}\sim\frac{1}{\gamma_2}, t_{\text{min}2}\sim\frac{\gamma_1}{J^2}$
      & $t_{\text{p}} \sim \gamma_2/J^2$,  \\
    \hline
    \end{tabular}%

\caption{Summary of the generalized Loschmidt echo dynamics in the weak and strong dissipation regimes. For the strong dissipation regime, the dynamics are further classified based on whether the ground state of $H_d$ is degenerate or non-degenerate (non-deg).}
  \label{tab:addlabel}
\end{table}

\subsection{Weak dissipation regime}

\begin{figure}[t] 
		\centering \includegraphics[width=0.7\textwidth]{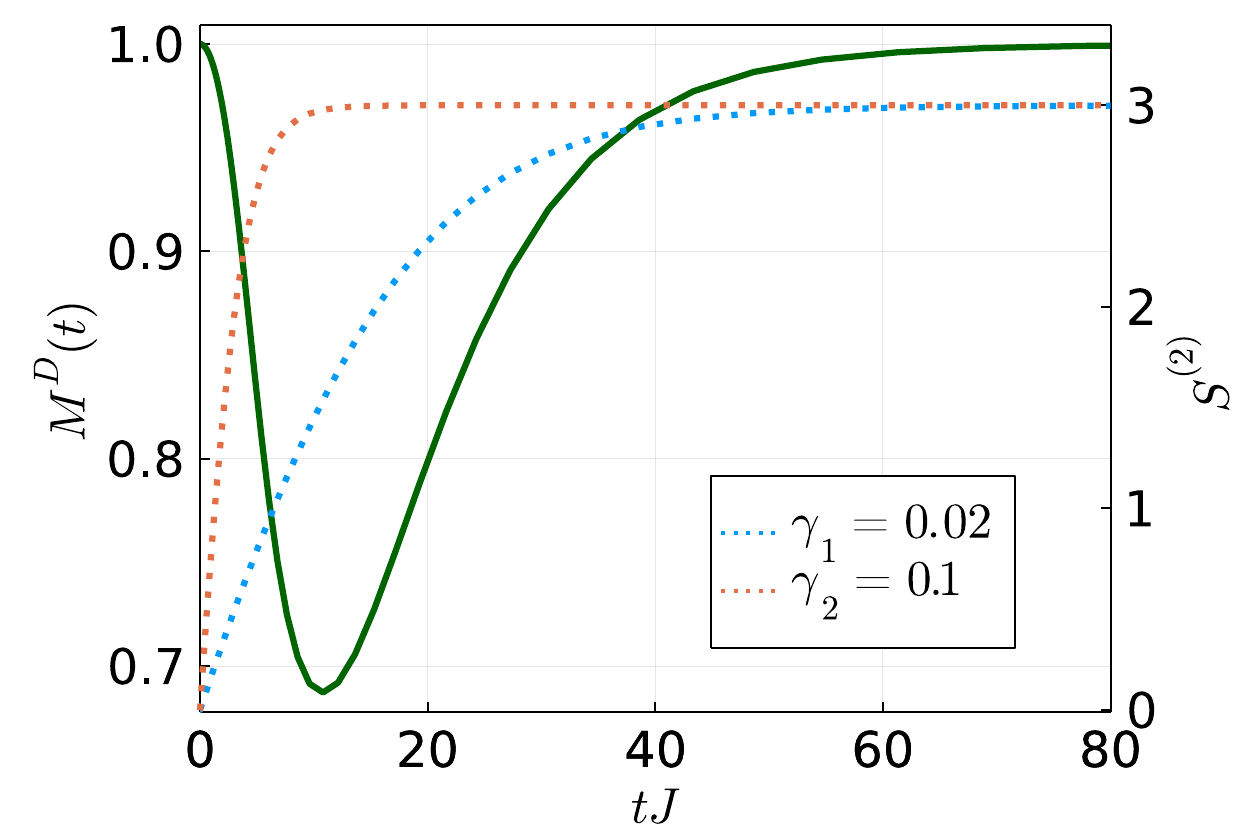} 
		\caption{The Loschmidt echo dynamics of the dissipative SYK model in the weak dissipation regime. We choose $N = 6$, $\gamma_1/J = 0.02$, and $\gamma_2/J = 0.1$, with dissipation applied to all Majorana fermions in the SYK model. The second R\'enyi entropy for the $\gamma_1$ and $\gamma_2$ evolutions are also plotted as blue and red dotted lines, respectively, for comparison. The initial state is the ground state of the SYK model. The time at which the Loschmidt echo reaches its minimum approximately coincides with the time when the second R\'enyi entropy for the smaller dissipation strength, $\gamma_1$, saturates to its final plateau value.}
		\label{echo_small_compare}
	\end{figure}

In the weak dissipation regime, where $\gamma_{1,2}/J \ll 1$, the generalized LE exhibits a universal behavior, as illustrated in Fig.~\ref{echo_small_compare}. The LE decays from its initial value of 1, reaching a minimum at the characteristic time scale $t_{\text{min}}$. Subsequently, the LE increases and eventually returns to a plateau with a value of 1 at the time scale $t_{\text{p}}$.

  For simplicity, we consider the case where the system has only one steady state, which is $\rho(t \to \infty) \propto \mathcal{I}$. Therefore, in the long-time limit, different dissipation strengths have no effect on the final states: both $\rho_1(t)$ and $\rho_2(t)$ approach the same steady state, and the LE returns to its initial value of 1.

The characteristic time scales $t_{\text{min}}$ and $t_{\text{p}}$ in the weak dissipation regime are determined by the dissipation strengths. To understand these time scales, we can analyze the dynamics of the second R\'enyi entropy, which is defined as follows:
    \begin{equation}
        S^{(2)}=-\log\left[\Tr (\rho^2)\right].
    \end{equation}
  The dynamics of the second R\'enyi entropy for $\rho_1$ and $\rho_2$ are shown in Fig.~\ref{echo_small_compare} by the blue and red dotted lines, respectively. For comparison, the green line represents the dynamics of the LE.

In the weak dissipation regime, the dynamics of the second R\'enyi entropy initially increase, following a characteristic $\gamma t$ scaling at short times \cite{add}, and eventually saturate as the initial state evolves toward the steady state. Since $\gamma_1 < \gamma_2$, the time for $\rho_2(t)$ to reach its steady state is shorter than that of $\rho_1(t)$. The time $t_{\text{min}}$ represents the time when one quantum state (under $\gamma_2$ evolution) reaches its steady state and its entropy stops increasing, while the other state (under $\gamma_1$ evolution) has not yet reached its steady state. Therefore, at $t_{\text{min}} \sim 1/\gamma_2$, the LE shows a minimum, as shown in Fig.~\ref{echo_small_compare}.

The final plateau time $t_{\text{p}} \sim 1/\gamma_1$ is the time when both $\rho_1(t)$ and $\rho_2(t)$ reach the same steady state, at which point both of their entropies reach their maximum values, as depicted in Fig.~\ref{echo_small_compare}, where the blue and red dotted lines saturate to the final plateau value.

\subsection{Strong dissipation regime}

The behavior of the LE becomes more complex when we consider the strong dissipation regime, where $\gamma_{1,2}/J \gg 1$. To analyze this, we use the double-space representation, where the time evolution of the density matrix is governed by the non-Hermitian double-space Hamiltonian $H^D = H_s - iH_d$, as defined in Eq.~\eqref{H_double}.

In the strong dissipation regime, $H^D$ is dominated by its dissipation part, $-i H_d$, while its Hermitian part, $H_s$, can be treated as a perturbation. Under these conditions, the dynamics of the generalized LE vary significantly depending on whether the ground state of $H_d$ is degenerate.

When the ground state of $H_d$ is non-degenerate, the LE dynamics follow patterns similar to those seen in the weak dissipation regime. However, if $H_d$ has degenerate ground states, we observe distinct behavior in the LE dynamics, including a potential two-local-minima structure, which can be attributed to the spectral structure of $H^D$ in this case.

\subsubsection{Ground state of $H_d$ is non-degenerate}
First, we consider the case where the dissipative Hamiltonian $H_d$ has a non-degenerate ground state. For example, we examine the dissipative SYK model in Eq.~\eqref{dissSYK_all_chain}, where dissipation acts on all Majorana fermions, i.e., $L_j = \chi_j$ for $j = 1, \dots, N$. In this model, the unique ground state of the dissipative Hamiltonian $H_d$ is the EPR state in double space, which corresponds to the identity density matrix in the original Hilbert space.

In our numerical simulation, we choose the initial state as the ground state of the SYK Hamiltonian. In this scenario, the LE dynamics, as shown in Fig.~\ref{echo_large_2}(a), exhibit a structure similar to that observed in the weak dissipation regime. This is because the typical behavior is still primarily determined by dissipation dynamics, as in the weak dissipation regime: at time $t_{\text{min}} \sim 1/\gamma_2$, $\rho_2$ with dissipation strength $\gamma_2$ nearly reaches its steady state, while the evolution under $\gamma_1$ still has non-vanishing dissipative components, causing the LE to reach a minimum. Then, at time $t_p \sim 1/\gamma_1$, both density matrices, $\rho_1$ and $\rho_2$, reach their steady states, and the LE saturates to a plateau.

\subsubsection{Ground state of $H_d$ is degenerate}
Next, we consider the case where the dissipative Hamiltonian $H_d$ has degenerate ground states. As an example, we examine a deformation of the dissipative SYK model, where only half of the Majorana fermions (with indices $j = 1, 2, \dots, \frac{N}{2}$) in the system are coupled to the environment. Under these conditions, the system's density matrix evolves as follows:
\begin{equation} 
		\frac{\partial{\rho}}{\partial t}=-i[H_{\text{SYK}},\rho] +2\gamma\sum_{j=1}^{N/2} \chi_j \rho \chi_j-\gamma\sum_{j=1}^{N/2}\lbrace \chi_j \chi_j, \rho\rbrace.\label{dissSYK_half_chain}
	\end{equation}

The ground state degeneracy of $H_d$ arises because dissipation only affects the degrees of freedom in one half of the system (Majorana fermions $\chi_j$ for $j = 1, 2, \dots, \frac{N}{2}$), leaving the energy of $H_d$ invariant in the other half. As a result, any pairing operator involving an even number of these undissipated Majorana fermions, such as $\chi_{j_1}\chi_{j_2}$ (with $j_1, j_2 = \frac{N}{2} + 1, \dots, N$), corresponds to a ground state of $H_d$.

\begin{figure*}[ht] 
\centering
\includegraphics[width=1.0\textwidth]{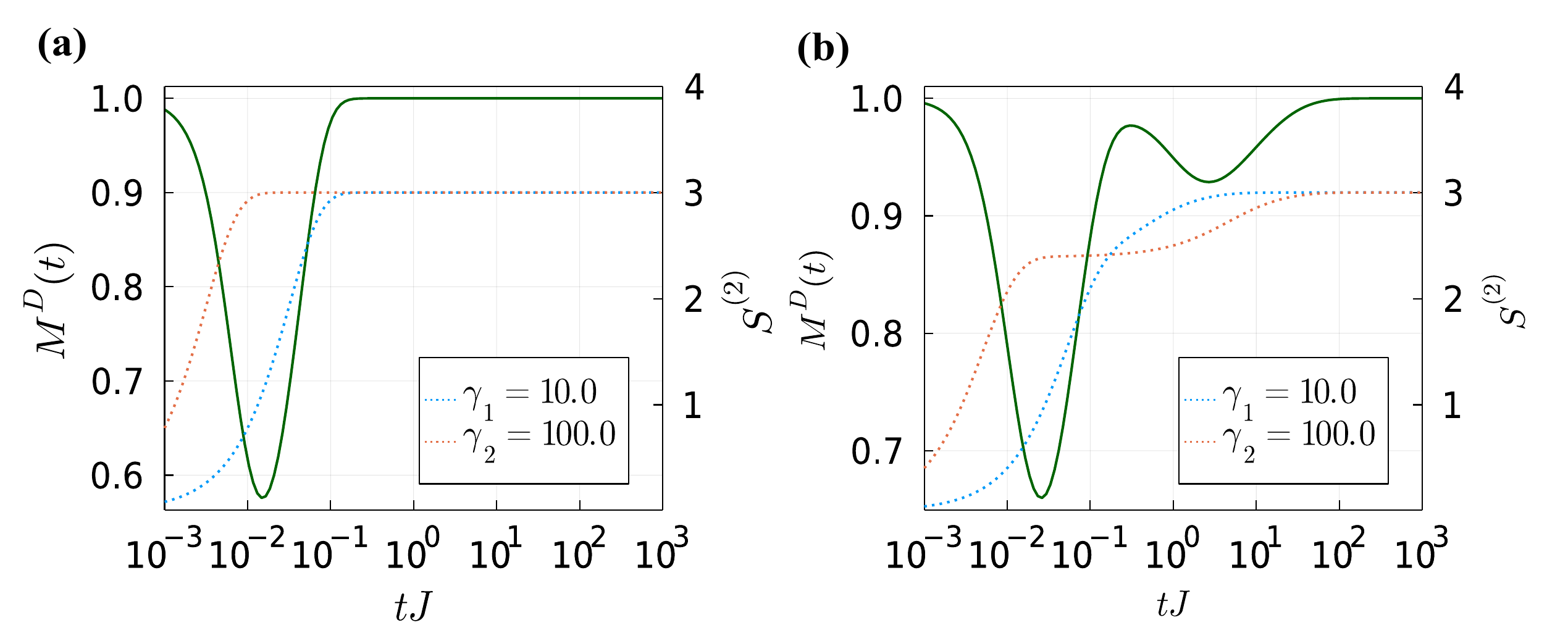} 
\caption{The Loschmidt echo dynamics of the dissipative SYK model in the strong dissipation regime. Here, we choose $N=6$, $\gamma_1/J=10$, and $\gamma_2/J=100$, with the initial state being the ground state of the SYK model. In panel (a), dissipation is applied to all Majorana fermions, resulting in a non-degenerate ground state of $H_d$,  while in panel (b), dissipation is applied to half of the Majorana fermions, leading to a degenerate ground state of $H_d$. The second R\'enyi entropy for $\gamma_1$ and $\gamma_2$ evolution are also plotted as the blue and red dot lines for comparison. The LE in (a) exhibits a single minimum structure, while the LE in (b) shows a two-local-minima structure.}

\label{echo_large_2}
\end{figure*}

In our numerical simulation, we choose the initial state as the ground state of the SYK Hamiltonian. This initial state contains components outside the ground-state Hilbert space of $H_d$, as the SYK Hamiltonian does not commute with $H_d$. Consequently, we observe that the generalized LE exhibits distinct dynamical behavior, characterized by a two-local-minima structure, as shown in Fig.~\ref{echo_large_2}(b).

\begin{figure}[t] 
	\centering 
	\includegraphics[width=0.7\textwidth]{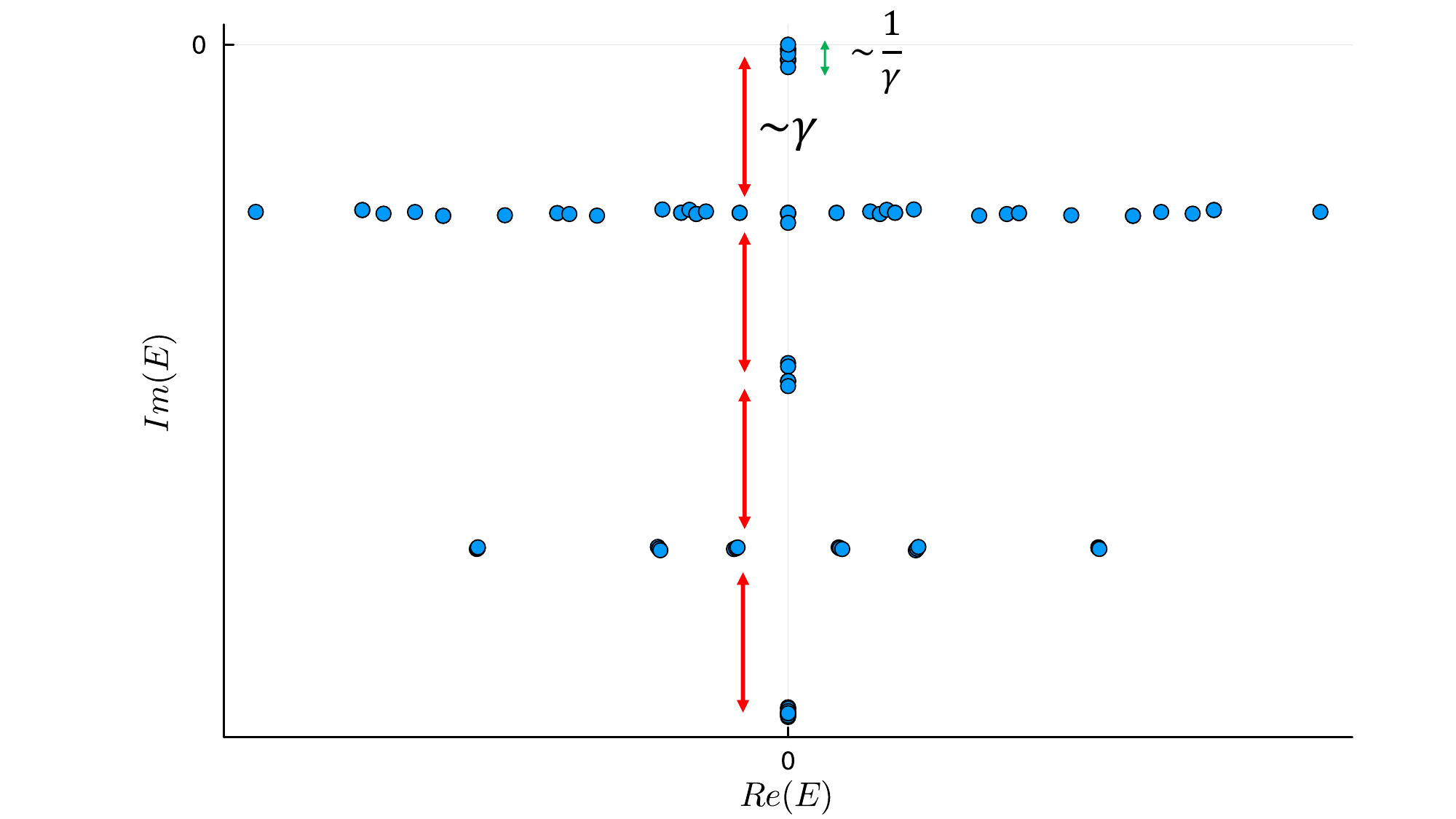} 
	\caption{The schematic diagram illustrates the spectrum of $H_D$ in the strong dissipation regime, where the spectrum splits into segments along the imaginary axis, spaced by intervals on the order of $\gamma$. The ground state degeneracy of $H_D$ is indicated by the presence or absence of a border around the segment near zero, with this border having a width on the order of $J^2/\gamma$. The figure shows the case where the ground states of $H_D$ are degenerate, highlighted by the bordered segment near zero. In contrast, if the ground state is non-degenerate, the segment around zero appears without this border.
 }
	\label{Lindblad_spectrum_cartoon}
\end{figure}

This novel structure is determined by the spectrum of $H_D$, which corresponds to the spectrum of the Lindblad superoperator and is referred to as the Lindblad spectrum. In the strong dissipation regime, the Lindblad spectrum exhibits two distinct energy scales, as illustrated in Fig.~\ref{Lindblad_spectrum_cartoon}. This separation of energy scales leads to contrasting behaviors in the short-time and long-time dynamics: as the dissipation strength increases, the short-time dynamics accelerate, while the long-time dynamics slow down.

More explicitly, the main feature of the Lindblad spectrum in the strong dissipation regime is that it separates into segments along the imaginary axis. Since $H^D$ is dominated by its dissipation part $-iH_d$ with dissipation strength $\gamma$, the separations between these segments are on the order of $\gamma$. Within each segment, the perturbative term $H_s$ in the double space Hamiltonian $H^D$ with energy scale $J$ determines the finer structure, resulting in widths along the imaginary axis on the order of $J^2/\gamma$, as depicted in Fig.~\ref{Lindblad_spectrum_cartoon}. We refer to the eigenstates with imaginary parts on the order of a few times $\gamma$ as high imaginary energy states, and the eigenstates with imaginary parts on the order of $J^2/\gamma$ as low-lying imaginary energy states \cite{zhouEnyiEntropyDynamics2021}.

For a general Lindblad evolution with this segmented Lindblad spectrum structure, the short-time dynamics are dominated by high-imaginary-energy states, which have imaginary components on the order of $\gamma$, leading to dynamics on the $\gamma t$ scale. In the long-time limit, these high-imaginary-energy states decay, and the dynamics become dominated by low-lying imaginary-energy states, with imaginary components on the order of $J^2/\gamma$. As a result, the long-time dynamics follow the $J^2t/\gamma$ scale.

With the above analysis of the Lindblad spectrum, we can now discuss the LE dynamics in the strong dissipation regime. Recall that we consider the case where the dissipation for both forward and backward evolutions ($\gamma_1$ and $\gamma_2$ evolution) has the same form, differing only in dissipation strength. As a result, the eigenstates of the double-space Hamiltonian for the forward and backward evolutions are nearly identical. Consequently, as we will demonstrate, the generalized LE exhibits a local maximum at intermediate times and a long-time plateau at a value of 1, due to the near alignment of eigenstates in the $\gamma_1$ and $\gamma_2$ evolutions.

The times at which the first and second local minima of the generalized LE occur are denoted as $t_{\text{min1}}$ and $t_{\text{min2}}$ respectively, with the local maximum between them marked as $t_{\text{max}}$. The time at which the LE reaches its final plateau is denoted as $t_{\text{p}}$. Below, we analyze the events occurring at each of these time points in chronological order and provide a brief discussion of the timescales associated with them.

\begin{itemize}
        \item First local minimum: \textbf{$t_{\text{min}1}$}.
         \\The initial decrease of the generalized LE is due to the differing decay rates of high imaginary energy states in the  $\gamma_1$ and $\gamma_2$ evolutions. The time $t_{\text{min}1}$ corresponds to the point when high imaginary energy states have mostly decayed in the $\gamma_2$ evolution but remain significant in the $\gamma_1$ evolution. Since the short-time dynamics follow the $\gamma t$ timescale, we have $t_{\text{min}1}\simeq \frac{1}{\gamma_2}$. After $t_{\text{min}1}$, the $\gamma_2$ evolution predominantly occupies low-lying imaginary energy states, while the high imaginary energy states in the $\gamma_1$ evolution, which are nearly orthogonal to the states in $\gamma_2$, begin to decay gradually. Consequently, the difference between the density matrices $\rho_1$ and $\rho_2$ decreases, leading to an increase in the LE after $t_{\text{min}1}$.

         \item Local maximum: \textbf{$t_{\text{max}}$}. 
        \\The local maximum occurs when the high-imaginary-energy states of the $\gamma_1$  evolution decay. At this point, both evolutions are predominantly distributed over low-lying imaginary energy states, resulting in the overlap between the density matrices $\rho_1$ and $\rho_2$ reaching a local maximum. Subsequently, the low-lying imaginary energy states begin to decay around $t\sim \frac{J^2}{\gamma}$ for the $\gamma_1$ evolution. As a result, the LE decreases again, since the decay rates of the low-lying imaginary energy states for $\gamma_1$ and $\gamma_2$ evolutions are different.

       \item Second local minimum: \textbf{$t_{\text{min}2}$}.
        \\At $t_{\text{min}2}$, the low-lying imaginary energy states have decayed for the $\gamma_1$ evolution, while they persist for the $\gamma_2$ evolution. At this time point, the $\gamma_1$ evolution approaches its steady state, whereas the $\gamma_2$ evolution has not yet reached a steady state. Given that the long-time dynamics scale as $\frac{J^2}{\gamma} t$ scale, the $t_{\text{min}2}\sim \frac{\gamma_1}{J^2}$. 

        \item Long-time plateau: \textbf{$t_{\text{p}}$}.
        \\Following $t_{\text{min}2}$,  the low-lying imaginary energy states for the $\gamma_2$ evolution gradually decay,  reaching steady state at  $t_{\text{p}} \sim \frac{\gamma_2}{J^2}$. At this point, the LE reaches its final plateau value of  $1$, indicating that both evolutions have reached their respective steady states.
    
    \end{itemize}

We compare the dynamics of the generalized LE with that of the second R\'enyi entropy for $\gamma_1$ and $\gamma_2$ evolutions in Fig.~\ref{echo_large_2}(b). The second R\'enyi entropy dynamics for $\rho_{1}$ and $\rho_2$ are shown by the blue and red dotted lines, respectively, while the LE dynamics are represented by the green line for comparison. As illustrated in Fig.~\ref{echo_large_2}(b), the timescale of the LE’s first minimum, $t_{\text{min}1}$, coincides with the point when the R\'enyi entropy dynamics of the $\gamma_2$  evolution reach their first plateau (means that the high imaginary energy states for $\gamma_2$ evolution decay out, while its low-lying imaginary energy states have not decayed yet). In contrast, the $\gamma_1$ evolution has not yet reached its first plateau, indicating that the high imaginary energy states are still present.

Similarly, the timescale of the LE's second minimum, $t_{\text{min}2}$, aligns with the maximum value of the R\'enyi entropy for the $\gamma_1$ evolution. This indicates that the low-lying imaginary energy states in $\gamma_1$ evolution have decayed, allowing it to reach a steady state, while the $\gamma_2$ evolution has not yet reached its maximum entropy. Finally, the time at which the LE reaches its final plateau coincides with the point at which the R\'enyi entropy dynamics of the $\gamma_2$  evolution reach their final saturation. After this point, both evolutions reach their steady states.

The LE behavior discussed in this paper is not specific to the SYK model but is quite general. In the weak-dissipation regime, the single-minimum structure of the LE follows directly from a perturbative analysis and does not rely on any special property of the SYK Hamiltonian. In the strong-dissipation regime, the emergence of two local minima in the open-system LE only requires that (i) the Hermitian part of the doubled-space Hamiltonian does not commute with the dissipative part, and (ii) the ground space of the dissipative part of the doubled-space Hamiltonian is degenerate (as is the case, for example, for generic local and translation-invariant Lindblad operators). These conditions are generic for open quantum systems and not specific to SYK. Consistently, numerical simulations of the XXZ model with dissipation, presented in Appendix C, show that its open-system LE dynamics exhibit the same structures.

\section{The OTOC-Loschmidt echo relation in open systems} \label{OTOC-LE_section}

The out-of-time-ordered correlator (OTOC) is a widely used diagnostic tool for detecting quantum chaos, as it measures quantum information scrambling and the system's sensitivity to initial conditions. In a closed system, it is defined as a four-point correlator with an unconventional time-ordering
\begin{equation}
	F_\beta(t)\equiv\langle  R_B^{\dagger}(t)W_A^{\dagger}R_B(t)W_A\rangle_\beta.
\label{OTOC_closed_def}
\end{equation} 
Here, $R_B(t)= e^{iHt}R_Be^{-iHt}$
represents the time evolution of the operator $ R_B $, $\langle \cdot \rangle_\beta$ denotes thermal average at temperature $1/\beta$. Below, we consider the infinite temperature case for simplicity. Thus, $F(t)\equiv \Tr \left[ R_B^{\dagger}(t)W_A^{\dagger}R_B(t)W_A\right]/d$,
where $d$ is the dimension of the Hilbert space.

Both OTOC and LE describe the irreversibility induced by perturbations, and they are connected through the OTOC-LE relation in closed systems~\cite{yanInformationScramblingLoschmidt2020}. In real systems, dissipation is inevitable, which raises the question: does this relation also hold in dissipative settings? In this section, we extend the definition of OTOC and establish a generalized OTOC-LE relation for open systems.

\subsection{OTOC-LE relation in closed systems}

We first review how OTOC and LE are related in closed systems.  Following the approach in~\cite{yanInformationScramblingLoschmidt2020},  consider a bipartite system where $A$ is a small subsystem and $B$ is its complement. We choose $ W_A $ and $ R_B $ in Eq.~\eqref{OTOC_closed_def} to be local unitary operators defined in subsystem $A$ and $B$, respectively. Since the dynamical behavior of the OTOC is universal for a chaotic Hamiltonian, which is insensitive to the specific choices of the operators $ W_A $ and $ R_B $, as long as they do not reflect the particular symmetries of the Hamiltonian, we can consider the average OTOC over all unitary operators supported on  subsystems $ A $ and $ B $,

\begin{equation}
	\overline{F(t)}=\frac{1}{d}\int dW_A dR_B  \Tr \left[ R_B^{\dagger}(t)W_A^{\dagger}R_B(t)W_A\right].
 \label{average_OTOC}
\end{equation} 
Here, the integration is performed with respect to the Haar measure for unitary operators. 

With the aid of the formula for the Haar integral
\begin{equation}
	\int dW_A W_A^{\dagger} O W_A = \frac{1}{d_A} \mathcal{I}_A\otimes \Tr_A(O),
 \label{Haar-integral}
\end{equation} 
the average  OTOC in Eq.~\eqref{average_OTOC} can be written as 
\begin{equation}
\begin{split}
	\overline{F(t)}&=\frac{1}{d_A d} \int dR_B \Tr \left\{\mathcal{I}_A\otimes \Tr_A\left[R_B(t)\right]R_B^{\dagger}(t)\right\}\\
    &=\frac{1}{d_A d}\int dR_B \Tr_B \left\{ \Tr_A\left[R_B(t)\right]\Tr_A\left[R_B^{\dagger}(t)\right] \right\}.
\end{split}
\end{equation} 
Thus, the OTOC is related to the reduced evolution of the operator $\text{Tr}_A \left[R_B(t) \right] = \text{Tr}_A \left( e^{iHt} R_B e^{-iHt} \right)$.

Without loss of generality, the total system Hamiltonian can be written as
\begin{equation}
H = H_A + H_B + V.
\end{equation}
 Here, $H_A$ and $H_B$ are the Hamiltonians defined on subsystems A and B, respectively. The interaction term is given by $V = \delta \sum_k V_A^k \otimes V_B^k$, where $\delta$ represents the coupling constant between subsystems A and B.
 To account for the effect of the coupling on the reduced evolution of the operator $R_B$, we approximately treat it as random noises acting on subsystem B~\cite{zurek2003decoherence,schlosshauer2007quantum,yanInformationScramblingLoschmidt2020},
\begin{equation}
	\Tr_A(e^{iHt}R_Be^{-iHt}) \simeq d_A\overline{e^{i(H_B +V_{\alpha})t}R_Be^{-i(H_B +V_{\alpha})t}}.
 \label{reduced_approx}
\end{equation} 
Here, $\{V_{\alpha}\}$ denotes the random noise operator, and the overline represents the average over all the realizations of $\{V_{\alpha}\}$. 
Here we use random external noise on subsystem $B$ to approximate the role of the interaction between the subsystem and its environment, a standard approach in studies of decoherence dynamics~\cite{zurek2003decoherence,schlosshauer2007quantum}. This approximation can be easily understood in the $\delta \ll 1$ limit~\cite{yanInformationScramblingLoschmidt2020}: the time evolution of an operator $R_B$ can be expanded to order $\delta^2$, leading to an effective master equation under the Born–Markov approximation. This master equation is equivalent to a stochastic time evolution of subsystem $B$ under a Langevin noise $\{V_\alpha\}$.

Using the above approximation and the Haar integral in Eq.~\eqref{Haar-integral} to compute the random average over $R_B$, the average OTOC in Eq.~\eqref{average_OTOC} can be further written as 
\begin{equation}
    \begin{aligned}
	\overline{F(t)} &\simeq \frac{1}{d_A d}\int dR_B \mathrm{Tr}_B \left\{ d_A\overline{e^{i(H_B +V_{\alpha})t}R_Be^{-i(H_B +V_{\alpha})t}} d_A\overline{e^{i(H_B +V_{\alpha^{'}})t}R_Be^{-i(H_B +V_{\alpha^{'}})t}} \right\}\\
    &=\frac{1}{d_B^2}\overline{|\mathrm{Tr}_B\left[e^{-i(H_B +V_{\alpha})t}e^{i(H_B +V_{\alpha^{'}})t}\right]|^2},
    \end{aligned} 
\end{equation}
where the overline in the R.H.S of the final step represents the average over all realizations of the noise operators $\{V_{\alpha}\}$  and $\{V_{\alpha^{'}}\}$. Each term in this average corresponds to the LE as defined in Eq.~\eqref{Echo_closed_def}. Thus, we obtain the OTOC-LE relation in closed systems. We also provide proof of this OTOC-LE relation using the diagram representation in Appendix A~\cite{SM}.

\subsection{Generalization of OTOC-LE relation to open systems}
We proceed by defining the OTOC in open systems and demonstrate that a generalized OTOC-LE relation also holds in open systems. Previous studies have proposed various definitions for the OTOC in open systems~\cite{syzranov2018out,zanardi2021information,swingle2018resilience,zhang2019information,yoshida2019disentangling,dominguez2021decoherence,xu2021thermofield,xu2019extreme,tuziemski2019out,PhysRevA.103.062214}. Here, we adopt the definition that is particularly suitable for future experimental realizations, such as those using NMR techniques. The experimental implementation using NMR will be discussed in detail in Section \ref{OTOC_experiment}.

The definition of the OTOC in open systems is extended as follows:
\begin{equation}
\label{OTOC_def_open}
F^D(t) = \frac{1}{d} \Tr \left\{ R_B^{\dagger} e^{\mathcal{L}^{\dagger}t} \left[W_A^{\dagger} e^{\mathcal{L}t}[R_B] W_A\right] \right\}.
\end{equation}
We denote the OTOC of the open system as $F^D$, distinguishing it from the OTOC of closed systems. Here $e^{\mathcal{L}t}[R]$ represents the dynamical evolution of an operator $R$ under the operator Lindblad equation:
\begin{equation}
\frac{\partial R}{\partial t} = i[H, R] + 2\gamma \sum_m L_m^{\dagger} R L_m  - \gamma \sum_m \left\{ L_m^{\dagger} L_m, R \right\}
\label{operator_Lindblad_eq}
\end{equation}
for a total time $t$. Note that a minus sign should precede the $\sum_m L_m^{\dagger} R L_m$ term when both $R$ and $L_m$ are fermionic operators~\cite{PhysRevResearch.5.033085,schwarz2016lindblad}.

The term $e^{\mathcal{L}^{\dagger}t}[R]$ represents the backward evolution of an operator $R$ under the operator Lindblad equation:
\begin{equation} 
	\frac{\partial{R}}{\partial t}=-i[H,R] +2\gamma\sum_m L_m ^{\dagger} RL_m-\gamma\sum_m\lbrace L^{\dagger}_m L_m, R\rbrace.
 \label{back_Lindblad}
\end{equation}
We also define $e^{\tilde{\mathcal{L}}t}[R]$ through the relation $$\Tr\left[ R e^{\mathcal{L}t}[W]\right] = \Tr\left[ e^{\tilde{\mathcal{L}}t}[R]W\right],$$ which represents the adjoint dynamical evolution of an operator $R$ under the equation:
\begin{equation}
\frac{\partial{R}}{\partial t} = -i[H, R] + 2\gamma \sum_m L_m R L_m^{\dagger} - \gamma \sum_m \lbrace L_m^{\dagger} L_m, R \rbrace.
\label{adjoint_Lindblad}
\end{equation}
When the system and bath are decoupled, meaning the dissipation strength in the Lindblad evolution is set to zero, the OTOC definition reduces to that of closed systems, as given in Eq.~\eqref{OTOC_closed_def}. Additionally, if all jump operators are Hermitian, we have $e^{\tilde{\mathcal{L}}^{\dagger}t}[R] = e^{\mathcal{L}t}[R]$.

Physically, the open-system OTOC $F^D(t)$ captures the combined effects of unitary information scrambling and environmental decoherence. For a chaotic Hamiltonian subject to dissipation, its early-time decay is governed by both the scrambling dynamics and the dissipative channels, and thus encodes their competition. A systematic analysis of how this decay decomposes into `chaotic' and `dissipative' contributions, and of its scaling behavior, would be an interesting problem in its own right.

We consider the average OTOC in open systems by averaging over unitary operators on subsystems $A$ and $B$:
\begin{equation}
	\overline{F^D(t)}=\frac{1}{d}\int dW_A dR_B  \Tr\lbrace R_B^{\dagger} e^{\mathcal{L}^{\dagger}t}\left[W_A^{\dagger}e^{\mathcal{L}t}[R_B]
 W_A\right]   \rbrace.
 \label{average_open_OTOC}
\end{equation} 
 Here, $W_A$ and $R_B$ represent random unitary operators acting on subsystems A and B, respectively. Averaging the OTOC over the random unitary operators $W_A$ yields
\begin{equation}
	\begin{split}
		\int dW_A F^D(t)
        =&\frac{1}{d_A}\Tr\left\{ \mathcal{I}_A \otimes \Tr_A \left[  e^{\mathcal{L}t}\left[R_B\right]\right]  e^{\tilde{\mathcal{L}^{\dagger}}t}\left[R_B^{\dagger}\right] \right\} \\
		=&\frac{1}{d_A}\Tr_B\left\{ \Tr_A \left[  e^{\mathcal{L}t}[R_B]\right] \Tr_A \left[ e^{\tilde{\mathcal{L}^{\dagger}}t}[R_B^{\dagger}] \right] \right\}.\\
	\end{split}
\end{equation} 
We then apply an approximation similar to that used in closed systems~\cite{open_approx_comment}, where the effect of the coupling on the reduced evolution of the operator $R_B$ is approximated as random noise, represented by ${V_{\alpha}}$, acting on subsystem B. This approximation leads to the expression: $$ \Tr_A \left[  e^{\mathcal{L}t}[R_B]\right] \simeq d_A\overline{e^{\mathcal{L}_{B,\alpha}t}[R_B]},$$ the average OTOC further becomes
\begin{equation}
	\begin{split}
        \overline{F(t)}\simeq &\frac{1}{d_B}\int dR_B \overline{\Tr_B \lbrace e^{\mathcal{L}_{B,\alpha} t}[R_B]e^{\tilde{\mathcal{L}}_{B,\alpha^{'}}^{\dagger}t}[R_B^{\dagger}] \rbrace} \\
	\end{split}
 \label{average_OTOC_open}
\end{equation} 
Here, $\alpha$ labels different realizations of the noise. $e^{\mathcal{L}_{B,\alpha}t}[R_B]$ represents the effective Lindblad evolution of operator $R_B$ under the Lindblad evolution
\begin{equation} 
	\frac{\partial{R}}{\partial t}=i[H+V_{\alpha},R] +2\gamma\sum_m L_m ^{\dagger} RL_m-\gamma\sum_m\lbrace L^{\dagger}_m L_m, R\rbrace.
\end{equation}
For simplicity, we have omitted the subscript $B$ in the above equation. The overline in $\eqref{average_OTOC_open}$ denotes the average over the random realizations of the noise operators $\{V_{\alpha}\}$ and $\{V_{\alpha'}\}$ acting on subsystem $B$.

For a general bipartite open system governed by the Lindblad evolution, we apply the Choi-Jamiolkowski isomorphism to map the evolution into a doubled Hilbert space. In this representation, the Hamiltonian in the doubled space can be formulated as:
\begin{equation*} 
	H^D=H^D_A + H_B^D  + V^D.
\end{equation*}
Here,
\begin{equation*} 
	H_{A/B}^D= H_{s,A/B}+H_{d,A/B},
\end{equation*}
and
\begin{equation*} 
	V^D = V_L\otimes \mathcal{I}_R - \mathcal{I}_L\otimes V^T_R,
\end{equation*}
where $H_{s,A/B}$ and $H_{d,A/B}$ are defined as in Eq.~\eqref{H_double}.

After mapping to the doubled space, the average OTOC can be expressed as 
\begin{equation}
	\begin{split}
		\overline{F^D(t)} 
  =& \frac{1}{d_B}\int dR_B \overline{\Tr_B \lbrace e^{-i(H^D_B+V_{\alpha}^D)t}  |R_B\rangle \langle R_B| e^{i(H^{D \dagger}_B+V_{\alpha^{'}}^{D\dagger})t}\rbrace} \\		
        =&\frac{1}{d_B^2}\overline{\Tr_B \left[ e^{i(H^{D \dagger}_B+V_{\alpha}^{D\dagger})t}  e^{-i(H^D_B+V_{\alpha^{'}}^D)t}\right]}.\\
	\end{split}
  \label{OTOC_open}
\end{equation} 
We then compare the above results with Eq.~\eqref{Loschmidt_open_def} and find that the averaged OTOC corresponds to the thermal average of (unnormalized) LE in open systems. This establishes a generalized relation between two distinct measures of information scrambling—OTOC and LE—applicable to open quantum systems governed by Lindblad time evolution. The connection between the OTOC and LE offers valuable insights into their dynamics in open systems, particularly in understanding quantum chaos within dissipative environments. A diagrammatic proof of this OTOC-LE relation in open systems is given in Appendix B~\cite{SM}.

This OTOC--LE relation in open systems is not unique: in the presence of dissipation there is no canonical definition of either quantity, so our construction should be viewed as one consistent framework rather than a mathematically unique one. Our derivation has two main steps. First, we average over operators $W_A$ on subsystem $A$, allowing us to rewrite the OTOC in terms of the reduced time evolution of an operator $R_B$ on subsystem $B$, so the LE--OTOC relation is tied to this averaging. Second, we model the reduced dynamics of $R_B$ as evolution under an effective Hamiltonian on $B$ plus stochastic noise induced by the coupling between $A$ and $B$, so the resulting LE--OTOC connection holds within this random-noise framework. Finding a more direct and fully general relation between LE and OTOC in open systems, without these additional assumptions, remains an interesting open problem.

\section{The OTOC-R\'enyi Entropy Relation in Open Systems} 
\label{OTOC-RE_section}

In the following, we will demonstrate that the normalization factor for the LE in open systems, which is the purity of the density matrix, is also related to the average of the OTOC. We begin by reviewing the relation between the OTOC and R\'enyi entropy in the closed system (a diagrammatic proof of this OTOC-R\'enyi entropy relation can be found in \cite{fanOutofTimeOrderCorrelationManyBody2017}). We will then show that this relation holds in open systems as well.

We consider a system initialized at infinite temperature $\rho \propto \mathcal{I}$, then it is quenched by an arbitrary operator $O$ at time $t = 0$. Thus, $\rho(0) = O O^{\dagger}$. The density matrix at time $t$ is then given by $\rho(t) = U(t) O O^{\dagger} U^{\dagger}(t)$. $U(t)=e^{-iHt}$ is the time evolution operator. We divide the total system into subsystems A and B, the reduced density matrix of subsystem A is
\begin{equation}
\rho_A(t)=\Tr_B\left[ \rho(t)\right]=\Tr_B\left[ U(t)OO^{\dagger}U^{\dagger}(t)\right].
\end{equation}
We consider the second   R\'enyi entropy $S^{(2)}_A$ for the subsystem $A$ defined as
\begin{equation}
S_A^{(2)}=-\log\left[\Tr_A (\rho_A^2)\right].
\end{equation}
We consider the case of infinite temperature for simplicity. The OTOC is then defined as $$F(t) = \Tr \left[V^{\dagger}(t) R_B^{\dagger} V(t) R \right],$$ where we choose $V \equiv \rho(0) = O O^{\dagger}$. A general relation connecting the OTOC and the second R\'enyi entropy can be obtained by considering the average OTOC over the operator $R_B$  \cite{fanOutofTimeOrderCorrelationManyBody2017} 
\begin{equation}
    \begin{split}
        \int dR_B \Tr  \left[ V^{\dagger}(t)R_B^{\dagger}V(t)R_B  \right]
        &= \frac{1}{d_B}\Tr  \left[ \Tr_B[V(t)]\otimes \mathcal{I}_B V(t) \right]\\
        &= \Tr_A  \left[ \Tr_B[V(t)]\Tr_B[V(t)] \right]\\
        &=\Tr_A (\rho_A^2).
    \end{split}
\end{equation} 
Here, $R_B$ is a local unitary operator defined in subsystem B. The integral is performed over unitary operators with respect to the Haar measure. However, it is worth noting that the random operator $R_B$ only needs to satisfy the requirements of a unitary 1-design and does not necessarily need to be Haar random. In the second line of the equation above, we have applied the formula
\begin{equation}
	\int dR_B R_B^{\dagger}V R_B =\frac{1}{d_B}\Tr_B (V)\otimes I_B,
 \label{Haar_sum}
\end{equation} 
 and $V=V^{\dagger}=\rho(0)$. Thus, we obtain a general relation between the OTOC and the second R\'enyi entropy,
\begin{equation}
	\text{exp}(-S_A^{(2)})=\int dR_B \Tr  \left[ V^{\dagger}(t)R_B^{\dagger}V(t)R_B \right].
\end{equation}

Below, we show that this relation still holds in open systems:
\begin{equation}
	\text{exp}(-S_A^{(2)})=
    \int dR_B \Tr \left\{ V^{\dagger} e^{\mathcal{L}^{\dagger}t} \left[R_B^{\dagger} e^{\mathcal{L}t}[V] R_B\right] \right\},
 \label{open_RE_OTOC}
\end{equation}
where the Lindblad operators are assumed to be Hermitian, which implies $e^{\tilde{\mathcal{L}}^{\dagger}t}=e^{\mathcal{L}t}$.
The reduced density matrix of subsystem A of the open system is given by:
\begin{equation}
\rho_A(t)=\Tr_B\rho(t)=\Tr_B\left[ e^{\mathcal{L}t}[V]\right].
\end{equation}
Use the Eq.~\eqref{Haar_sum}, the R.H.S of the Eq.~\eqref{open_RE_OTOC} can be rewritten as
\begin{equation}
\begin{split}
    \int dR_B \Tr \left\{ V^{\dagger} e^{\mathcal{L}^{\dagger}t} \left[R_B^{\dagger} e^{\mathcal{L}t}[V] R_B\right] \right\}
    &= \frac{1}{d_B}\Tr\lbrace e^{\tilde{\mathcal{L}}^{\dagger}t}[V](\Tr_Be^{\mathcal{L}t}[V])\otimes I_B   \rbrace \\
    &=\Tr_A\left[ (\Tr_Be^{\tilde{\mathcal{L}}^{\dagger}t}[V]) (\Tr_Be^{\mathcal{L}t}[V]) \right] \\
    &=\Tr_A\left[ \rho_A^2 \right]\\
    &=\text{exp}(-S_A^{(2)}).
\end{split}
\end{equation}
The third line follows from $e^{\tilde{\mathcal{L}}^{\dagger}t}=e^{\mathcal{L}t}$. Additionally, we used $V^{\dagger}=V$ since $V=O O^{\dagger}$. This completes the proof of Eq.~\eqref{open_RE_OTOC}.

This relation, as given in Eq.~\eqref{open_RE_OTOC}, demonstrates that in open systems, the second R\'enyi entropy can be used to infer properties of the OTOC. The L.H.S. of the equation represents the R\'enyi entropy, which can be measured in a quench experiment, while the R.H.S. represents an equilibrium average with respect to the infinite-temperature density matrix. This establishes a connection between correlations evaluated in an equilibrium density matrix and quantities measured in non-equilibrium processes. Analogous to linear response theory—where a normal correlator measures the response of observables to a perturbation of the system's Hamiltonian, the OTOC measures the response of the system's entropy to a quench~\cite{fanOutofTimeOrderCorrelationManyBody2017}.

The second R\'enyi entropy in an open quantum system quantifies information lost from the system to its environment. For a pure state evolving under closed Hamiltonian dynamics, $S_2 = 0$, whereas coupling to a bath or performing measurements increases $S_2$ through decoherence and dissipation. The time dependence $S_2(t)$ can be fitted to extract dephasing rates, mixing times, and the Liouvillian gap. In many-body settings, the growth and saturation of subsystem $S_2$ reflect the competition between unitary entanglement generation and monitoring/dissipation, and serve as diagnostics of dissipative or measurement-induced phase transitions. Moreover, $S_2$ is directly accessible experimentally via SWAP \cite{Daley_2012, Abanin_2012, 2015Natur.528...77I, PhysRevX.6.041033} or randomized-measurement protocols on two copies of the system \cite{Elben_2018, 2019Sci...364..260B, Elben_2019}.

In this section and the previous one, we separately consider the non-normalized LE and the normalization factor of LE. The non-normalized LE is related to the average OTOC, while the normalization factor, which is directly related to the R\'enyi entropy, is also linked to the average OTOC. This approach aligns with real experimental conditions, where these quantities are more naturally measured independently and then combined to define the LE in the open system.

\section{The experiment protocol of measuring OTOC}	\label{OTOC_experiment}

In closed systems, the OTOC has been experimentally measured on various platforms, including NMR~\cite{PhysRevX.7.031011, PhysRevLett.120.070501}, superconducting circuits~\cite{PhysRevLett.125.120504, Braum_ller_2021, Mi_2021}, trapped ions~\cite{G_rttner_2017, Landsman_2019, Joshi_2020}, and cold atoms~\cite{Pegahan_2021}. Below, we outline a straightforward protocol for measuring the OTOC defined in Eq.~\eqref{OTOC_def_open} in open systems, using an NMR experimental setup as an example.

This  protocol involves the following steps:
\begin{itemize}
    \item \textbf{1}. Initial state preparation:
    Prepare the system in a high-temperature state $\rho_0$ with a polarization field $M_B$~\cite{PhysRevX.7.031011}:
\begin{equation}
\rho_0 \propto e^{-\beta (H - h M_B)} \simeq \mathcal{I} - \epsilon M_B.
\end{equation}
Here, $\mathcal{I}$ represents the identity operator. Since $\mathcal{I}$ remains constant and does not affect observables under Lindblad dynamics, we treat the density matrix as effectively proportional to $M_B$.

\item \textbf{2}. Forward Evolution: Evolve the system with forward Lindblad evolution, then the system's density matrix becomes $e^{\mathcal{L}t}[M_B]$.

\item \textbf{3}.  Apply Perturbation: Apply an operator $W_A$ to the system,  resulting in the modified density matrix: $\rho_1 = W_A^{\dagger} e^{\mathcal{L}t}[M_B] W_A$.

\item \textbf{4}. Backward Evolution: Evolve the system with backward Lindblad dynamics, yielding the final state $\rho_2 = e^{\mathcal{L}^{\dagger}t} \left[ W_A^{\dagger} e^{\mathcal{L}t}[M_B] W_A \right]$.

\item \textbf{5}. Measurement: Measure the operator $M_B$ on the final state $\rho_2$.

\end{itemize}
The expectation value $M_B$ on the final state provides the OTOC:
\begin{equation}
    \Tr \left[  \rho_2 M_B \right] = \Tr \left\{ M_B e^{\mathcal{L}^{\dagger}t} \left[W_A^{\dagger} e^{\mathcal{L}t}[M_B] W_A\right] \right\}.
\end{equation}

This expression matches the OTOC for an open system as defined in Eq.~\eqref{OTOC_def_open} if we set $R_B = M_B$, and assume $M_B=M_B^{\dagger}$. Consequently, this protocol can be employed to experimentally measure the OTOC in open systems.

The step $4$ of this measurement protocol requires a time-reversal step. In closed systems, this is implemented by flipping the sign of the Hamiltonian, as demonstrated experimentally \cite{PhysRevX.7.031011}. In our open-system setting described by a Lindblad equation, the corresponding time reversal is given by the conjugate Lindbladian evolution defined in Eq.\eqref{back_Lindblad}. For Hermitian jump operators, a direct comparison of Eq.\eqref{operator_Lindblad_eq} and Eq.\eqref{back_Lindblad} shows that the only change is $H \to -H$, while the dissipator remains unchanged. Thus, the experimental implementation is identical to the closed case: one simply reverses the sign of the Hamiltonian and leaves the dissipation unchanged. Also, the OTOCs have already been measured in several noisy, effectively dissipative platforms, from NMR \cite{PhysRevX.7.031011, PhysRevA.104.012402, Dom_nguez_2021} and trapped-ion experiments  \cite{Garttner:2016mqj, 2019Natur.567...61L} to noisy gate-based quantum processors \cite{Green_2022}.

\section{Conclusion} \label{summary}

In this paper, we developed a framework to generalize the Loschmidt echo (LE) to open quantum systems and analyzed its dynamics using the Choi–Jamiołkowski isomorphism, which maps quantum evolution onto a doubled Hilbert space. Within this framework, we related the LE dynamics to the structure of the Lindblad spectrum. In the weak-dissipation regime, the LE exhibits a single minimum, while in the strong-dissipation regime, it can develop two local minima, depending on whether the ground space of the dissipative part of the doubled-space Hamiltonian is degenerate.

The qualitative structure of the open-system LE in these two regimes is universal within our setting, as it follows from perturbative arguments rather than from any special property of the SYK Hamiltonian. In the weak-dissipation regime, the single-minimum structure arises directly from a perturbative expansion in the dissipation strength. In the strong-dissipation regime, the emergence of two local minima requires only that (i) the Hermitian part of the doubled-space Hamiltonian does not commute with the dissipative part, and (ii) the ground space of the dissipative generator is degenerate (as in generic local and translation-invariant Lindblad operators). These conditions are generic for open quantum systems and not specific to SYK. We substantiate this by combining analytic arguments with numerical simulations of both dissipative SYK and XXZ models, which consistently exhibit the same one-minimum (weak dissipation) and two–local-minima (strong dissipation) structures. While this work primarily considers differences in dissipation strength between forward and backward Lindblad evolutions, future investigations could examine cases involving variations in dissipation forms or Hamiltonian components. Such studies could reveal additional universal structures in LE dynamics.

Additionally, we extended the definition of the OTOC to open systems. Our proposed definition is well-suited for probing universal information-scrambling properties in open quantum systems and can be experimentally measured using techniques like NMR. We demonstrated that the unnormalized ensemble-averaged LE in an open system is directly related to the OTOC  averaged over all unitary operators on two subsystems, paralleling the well-known OTOC-LE relation in closed systems. 
Furthermore, for a broad class of density matrices, we prove that the relation between the second R\'enyi entropy and the corresponding OTOC continues to hold in open systems, establishing links between different measures of information scrambling.

Future research could further generalize the connections among various information-scrambling measures in open systems, such as the spectral form factor, relative entropy, and operator entanglement~\cite{Maldacena_2016, Kudler_Flam_2020}. These efforts would deepen our understanding of information dynamics in dissipative quantum systems and open new directions for both experimental and theoretical investigations.

Lastly, it would be interesting to explore the symmetry aspects of LE dynamics~\cite{Weidinger_2017}. For example, if the forward evolution respects certain symmetries while the backward evolution weakly breaks them, one could investigate how this weak symmetry breaking influences the LE dynamics. Our definition of the LE, which quantifies the similarity between two density matrices over time, could also be applied to study phase transitions during this process. In particular, it might serve as an order parameter for the transition from strong to weak symmetry breaking~\cite{ma2024topologicalphasesaveragesymmetries,lessa2024strongtoweakspontaneoussymmetrybreaking,sala2024spontaneousstrongsymmetrybreaking,xu2024averageexactmixedanomaliescompatible,huang2024hydrodynamicseffectivefieldtheory,Kuno_2024,gu2024spontaneous}, which can only occur in mixed states. Moreover, it would be interesting to investigate the symmetry properties of two Lindblad evolutions~\cite{PhysRevLett.124.040401,PhysRevX.13.031019,PRXQuantum.4.030328,Mao_2024} using our generalized LE shows promise for classifying the symmetry classes of different Lindbladians.

\section*{Acknowledgements}
We would like to thank Hui Zhai for introducing us to this interesting topic and for his many insightful discussions. We also thank Tian-Gang Zhou for his valuable feedback and advice in revising the manuscript.

\begin{appendix}
\numberwithin{equation}{section}

\section{The diagrammatic proof of the OTOC-LE relation in closed system}
We provide a diagrammatic proof of the OTOC-LE relation (the similar diagrammatic proof technique can be found in \cite{fanOutofTimeOrderCorrelationManyBody2017}), as reviewed in the Section \ref{OTOC-LE_section} of the main text. 

For a general operator $Q$, if we choose a complete orthogonal basis of subsystems A and B, we can write it as
\begin{equation}
	Q=\sum_{i_A,i_B,j_A,j_B}Q_{i_A,i_B,j_A,j_B}|i_A\rangle |i_B\rangle \langle j_A| \langle j_B|.
\end{equation} 

\begin{figure}[ht] 
		\centering \includegraphics[width=0.7\textwidth]{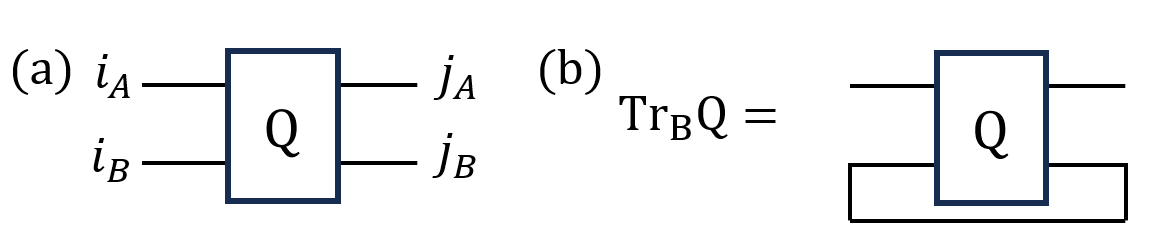} 
		\caption{The diagram represents the general operator $Q$ in (a) and the partial trace over subsystem B of it in (b).}
		\label{operator_trace_diagram}
	\end{figure}

The diagram illustrating this is shown in Fig.~\ref{operator_trace_diagram}(a), where the left legs represent the input and the right legs represent the output. When we perform a partial trace over the degrees of freedom of subsystem B, in the diagram, this involves connecting the input and output legs of subsystem B, as depicted in Fig.~\ref{operator_trace_diagram}(b). 

We will then use this diagrammatic technique to prove the OTOC-LE relation. For simplicity, we consider the infinite-temperature case, where $\rho(0)\propto \mathcal{I}$.

\begin{figure}[ht] 
		\centering \includegraphics[width=0.7\textwidth]{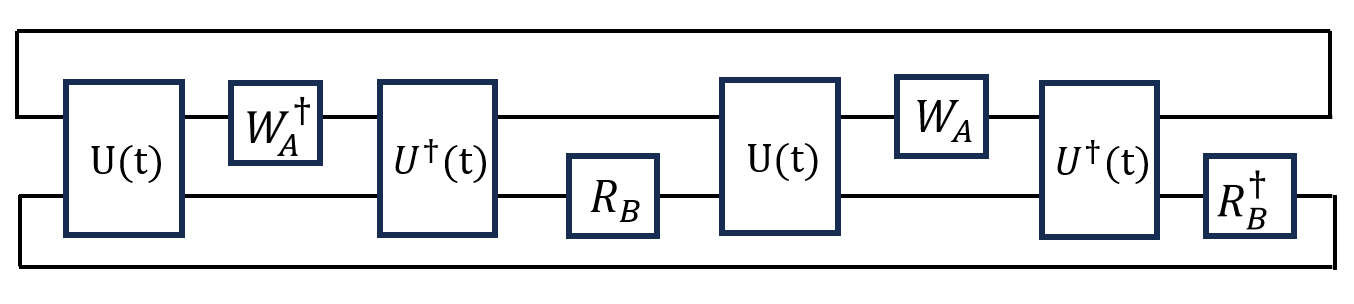} 
		\caption{The diagram representation of OTOC is defined in the Eq.~\eqref{OTOC_infinite_def}.}
		\label{otoc_diagram}
	\end{figure}

First, we can present the OTOC defined as 
\begin{equation}
	F_{\beta=0}(t)=\text{Tr}[ R_B^{\dagger}(t)W_A^{\dagger}R_B(t)W_A]
\label{OTOC_infinite_def}
\end{equation} 
in the  Fig.~\ref{otoc_diagram}.

\begin{figure}[ht] 
		\centering \includegraphics[width=0.7\textwidth]{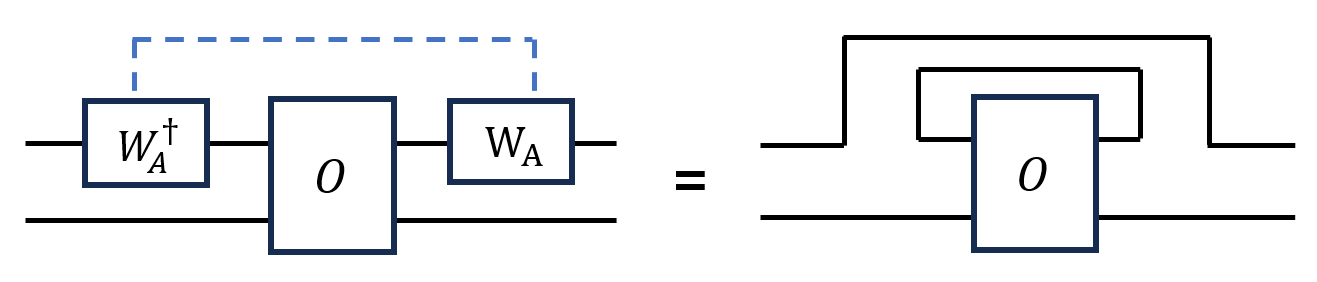} 
		\caption{The diagram representation of the  Haar random average formula Eq.~\eqref{Haar-random}. The blue dash line represents the Haar random average of the operator $W_A$ defined on subsystem A.}
		\label{Haar_average}
\end{figure}

Next, we perform the random averaging over operators on subsystem A
\begin{equation}
	\overline{F(t)}=\frac{1}{d_A}\int dW_A   \Tr \left[ R_B^{\dagger}(t)W_A^{\dagger}R_B(t)W_A\right].
 \label{average_OTOC_A}
\end{equation} 
We use the Haar random average formula,\begin{equation}
	\int dW_A W_A^{\dagger} O W_A = \frac{1}{d_A} \mathcal{I}_A\otimes \Tr_A(O),
 \label{Haar-random}
\end{equation} 
and it is depicted as the Fig.~\ref{Haar_average} (the constant $\frac{1}{d_A}$ has been omitted).

\begin{figure}[ht] 
		\centering \includegraphics[width=0.7\textwidth]{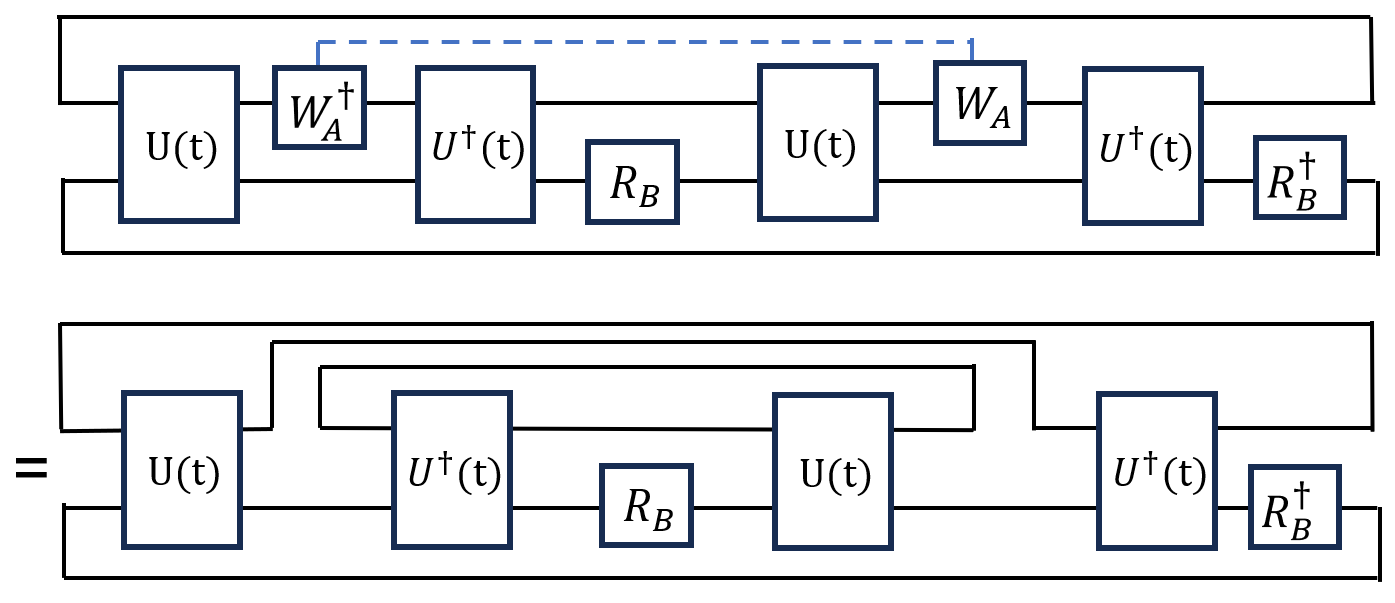} 
		\caption{The diagram representation of the average of OTOC over $W_A$.}
		\label{otoc_B_average}
\end{figure}

The diagrammatic representation of the average OTOC in Eq.~\eqref{average_OTOC_A} is shown in Fig.~\ref{otoc_B_average}, where the blue dash line represents the Haar random average of the operator $W_A$.

To account for the effect of the coupling on the reduced evolution of the operator $R_B$, we approximately treat it as random noises acting on subsystem B,
\begin{equation}
	\Tr_A(e^{iHt}R_Be^{-iHt}) \simeq d_A\overline{e^{i(H_B +V_{\alpha})t}R_Be^{-i(H_B +V_{\alpha})t}}.
\end{equation} 
The logic here is that the reduced evolution of $R_B$ can be viewed as a general open system dynamics where the open system (subsystem B) evolves together with the bath (subsystem A) unitarily. We consider the open system dynamics by tracing out the bath degree of freedom. Under the Born-Markov approximation, the open system evolution can be written as the form of the Lindblad master equation, where the form of the Lindblad jump operators is decided by the interaction between the system and the bath. The Lindblad evolution can be equivalently written as the system evolution under the effective Hamiltonian with random noise, this random noise has to satisfy some average properties, and this noise is called Langevin noise in literature.

\begin{figure}[ht] 
		\centering \includegraphics[width=0.45\textwidth]{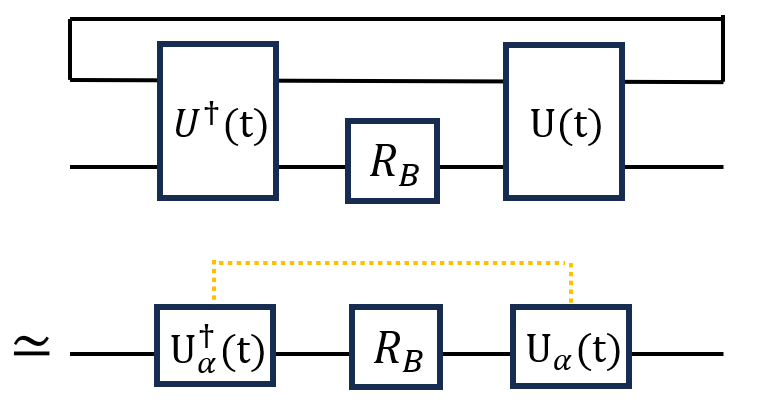} 
		\caption{The diagram representation of reduced evolution of $R_B$. Here, the effect of coupling on the reduced evolution of the operator $R_B$ is approximated as random noise $\{ V_{\alpha}\}$ on subsystem B, and the orange dot line denotes the random average over noise $\{ V_{\alpha}\}$.}
		\label{otoc_B_reduced}
\end{figure}

\begin{figure}[ht] 
		\centering \includegraphics[width=0.7\textwidth]{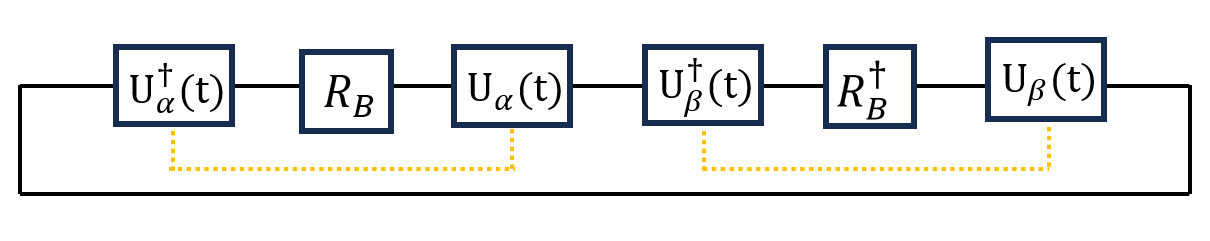} 
		\caption{The diagram representation of the average of OTOC over $W_A$ after using the approximation of Fig.~\ref{otoc_B_reduced}. Here, the orange dot line denotes the random average over noise on the subsystem B.}
		\label{reduced_OTOC}
\end{figure}

The diagrammatic representation of this approximation is shown in Fig.~\ref{otoc_B_reduced}, where the orange dotted line represents the random averaging over the noises $\{V_{\alpha}\}$.
Incorporating this diagram into the averaged OTOC, we obtain the diagram shown in Fig.~\ref{reduced_OTOC}.

\begin{figure}[ht] 
		\centering \includegraphics[width=0.7\textwidth]{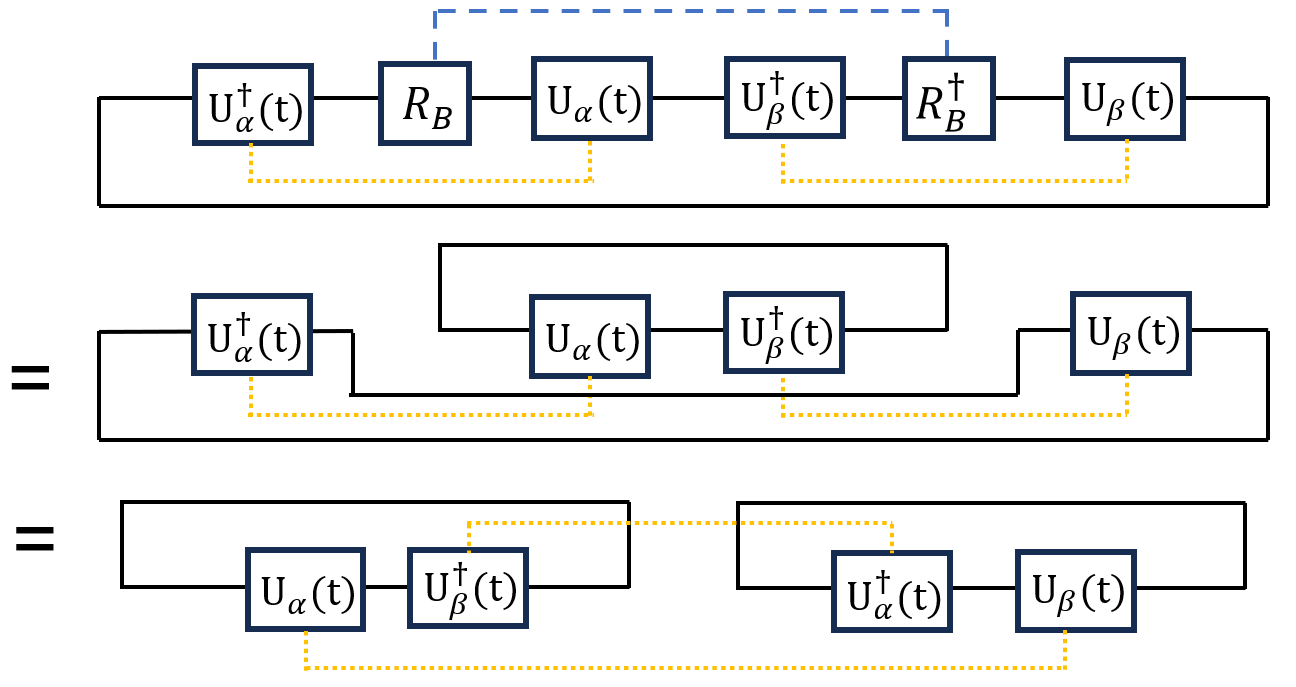} 
		\caption{The diagram representation of OTOC averaged over the random unitary operator on subsystems $A$ and $B$. Here, the blue dash line represents the random average over the unitary operators $R_B$, and the orange dot lines denote the random average over noise on subsystem B.}
		\label{otoc_AB_average}
\end{figure}

Next, we consider the averaging of operators on subsystem B, with the diagrammatic representation shown in Fig.\ref{otoc_AB_average}. Note that the final line in Fig.\ref{otoc_AB_average} takes the form of the LE.
Through these procedures, we have shown that the averaged OTOC in Eq.~\eqref{average_OTOC} can be further expressed as
\begin{equation}
	\overline{F(t)} \simeq \frac{1}{d_B^2}\overline{|\Tr_B\left[e^{-i(H_B +V_{\alpha})t}e^{i(H_B +V_{\beta})t}\right]|^2}.
\end{equation}
Recall that the definition of the LE at infinite temperature is 
\begin{equation}
		M(t)=|\text{Tr}[e^{iH_1 t} e^{-iH_2t}]|^2.
	\label{Echo_closed_infinite}
	\end{equation} 
We then observe that the final line of Fig.~\ref{otoc_AB_average} takes the form of the LE.

\section{The diagrammatic proof of the OTOC-LE relation in open system}
Below, we provide a diagrammatic proof of the OTOC-LE relation for open systems, which is quite similar to the proof for closed systems, except that the evolution is now governed by Lindblad dynamics rather than unitary evolution. For simplicity, we consider the infinite temperature case, where $\rho(0)\propto \mathcal{I}$. and we assume the Lindblad jump operators are all hermitian, thus $\tilde{\mathcal{L}}^{\dagger}=\mathcal{L}$ (the definition of $\tilde{\mathcal{L}}$ denote the backward Lindblad evolution and $\mathcal{L}^{\dagger}$ denote the adjoint Lindbald evolution, their definition are in the Eq.~\eqref{back_Lindblad} and Eq.~\eqref{adjoint_Lindblad} respectively).

\begin{figure}[ht] 
		\centering \includegraphics[width=0.7\textwidth]{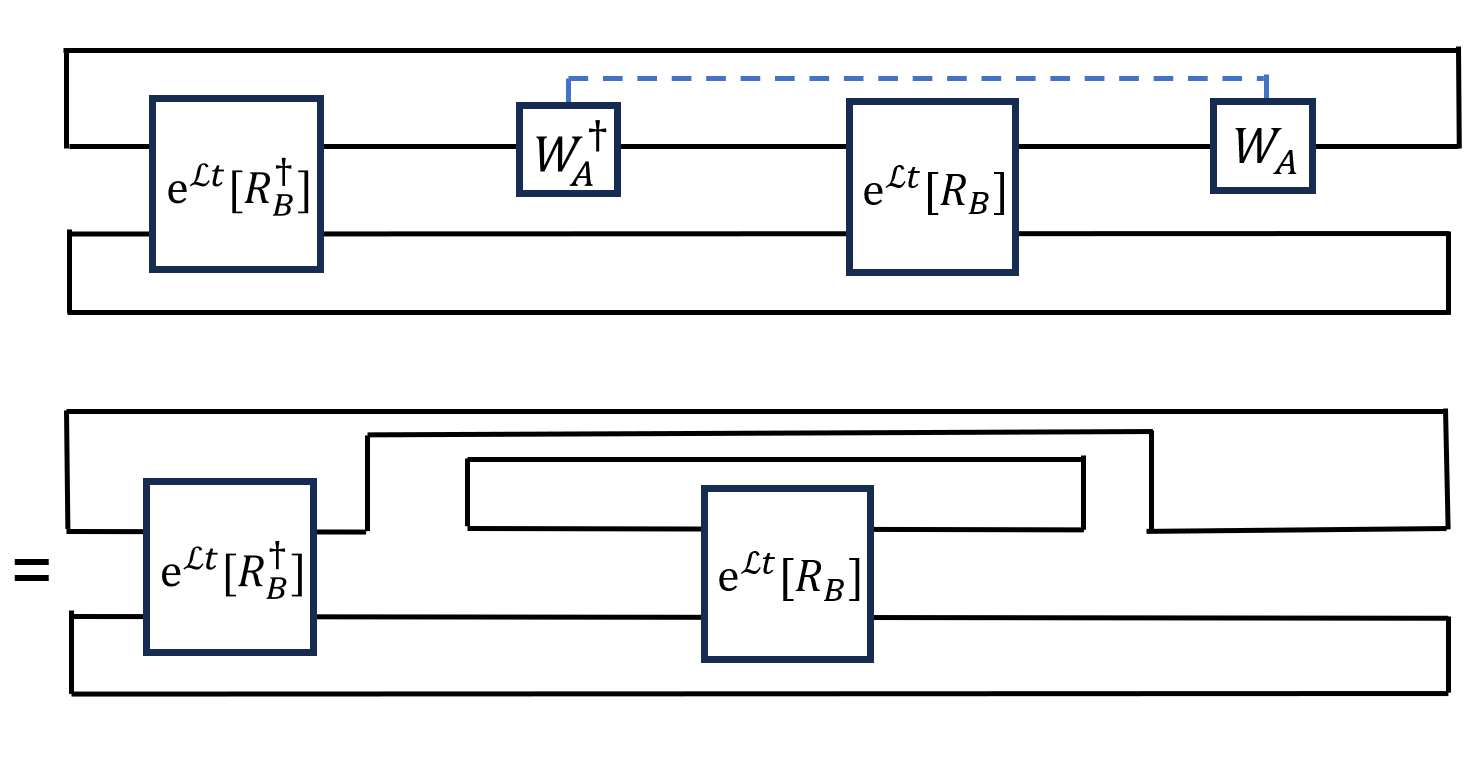} 
		\caption{The diagram representation of the average OTOC of operator $W_A$ supported on the subsystem A.}
		\label{open_otoc_B_average}
\end{figure}

 The average OTOC over the random unitary operator $W_A$ defined on subsystem A
\begin{equation}
	\begin{split}
		\int dW_A F^D(t)
		=\frac{1}{d_A}\int dW_A  \mathrm{Tr}\lbrace R_B^{\dagger} e^{\mathcal{L}^{\dagger}t}\left[W_A^{\dagger}e^{\mathcal{L}t}[R_B]
 W_A\right]   \rbrace.
	\end{split}
\end{equation} 
is represented in Fig.~\ref{open_otoc_B_average}.

\begin{figure}[ht] 
		\centering \includegraphics[width=0.5\textwidth]{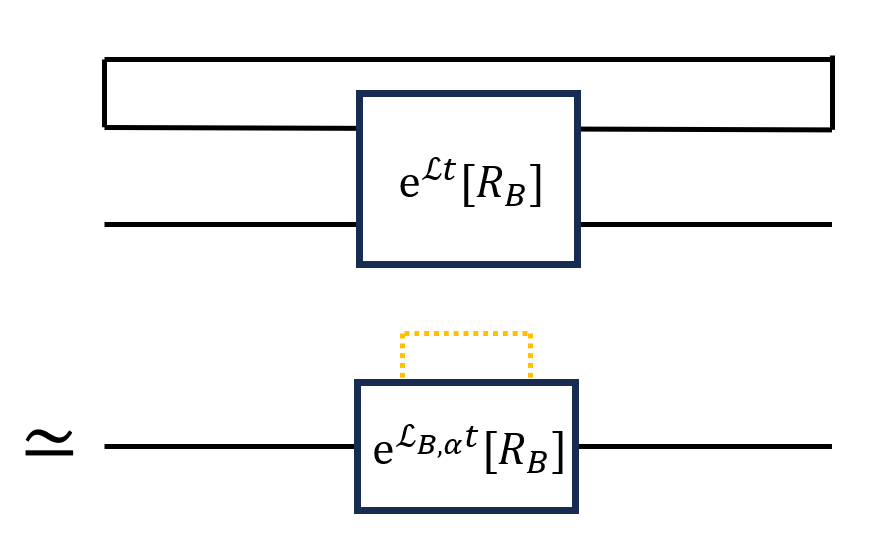} 
		\caption{The diagram representation of reduced evolution of $R_B$ in open system. Here, the effect of coupling on the reduced evolution of the operator $R_B$ is approximated as random noise $\{ V_{\alpha}\}$ on subsystem B, and the orange dot line denotes the random average over noise $\{ V_{\alpha}\}$.}
		\label{open_reduced_B}
\end{figure}

Here, we consider local dissipation, which does not involve interactions between subsystems A and B. Therefore, we use the same approximation as in the proof of the OTOC-LE relation for closed systems, where the effect of coupling on the reduced evolution of the operator $R_B$ is approximated as random noise on subsystem B.
This approximation can be generally formulated as $$ \Tr_A \left[  e^{\mathcal{L}t}[R_B]\right] \simeq d_A\overline{e^{\mathcal{L}_{B,\alpha}t}[R_B]},$$ and it is shown as Fig.~\ref{open_reduced_B}.

\begin{figure}[ht] 
		\centering \includegraphics[width=0.68\textwidth]{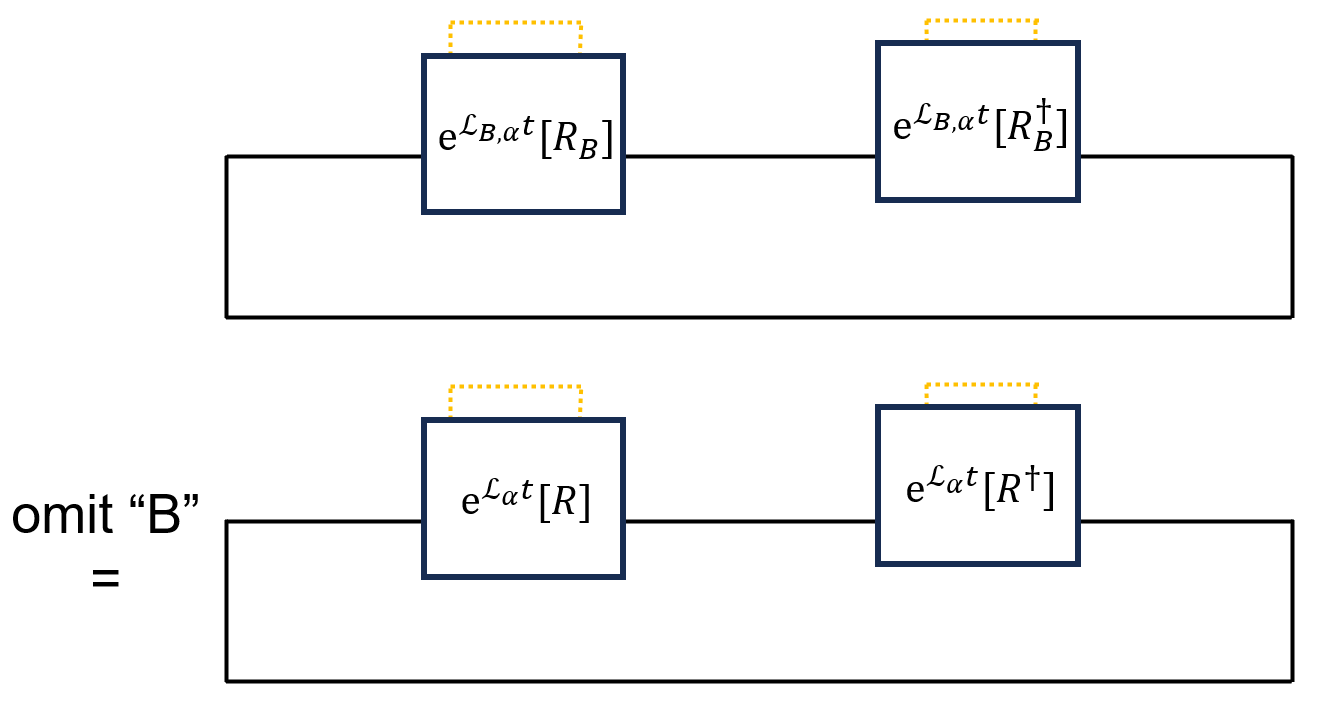} 
		\caption{The diagram representation of the average of OTOC after averaging over operators on subsystem A. We have omitted the subscript B since both the time evolution and operators only non-trivially act on subsystem B.}
		\label{omitB}
\end{figure}

Incorporating this diagram into the averaged OTOC, we obtain the diagram shown in Fig.~\ref{omitB}, where we omit the subscript B.

\begin{figure}[ht] 
		\centering \includegraphics[width=0.7\textwidth]{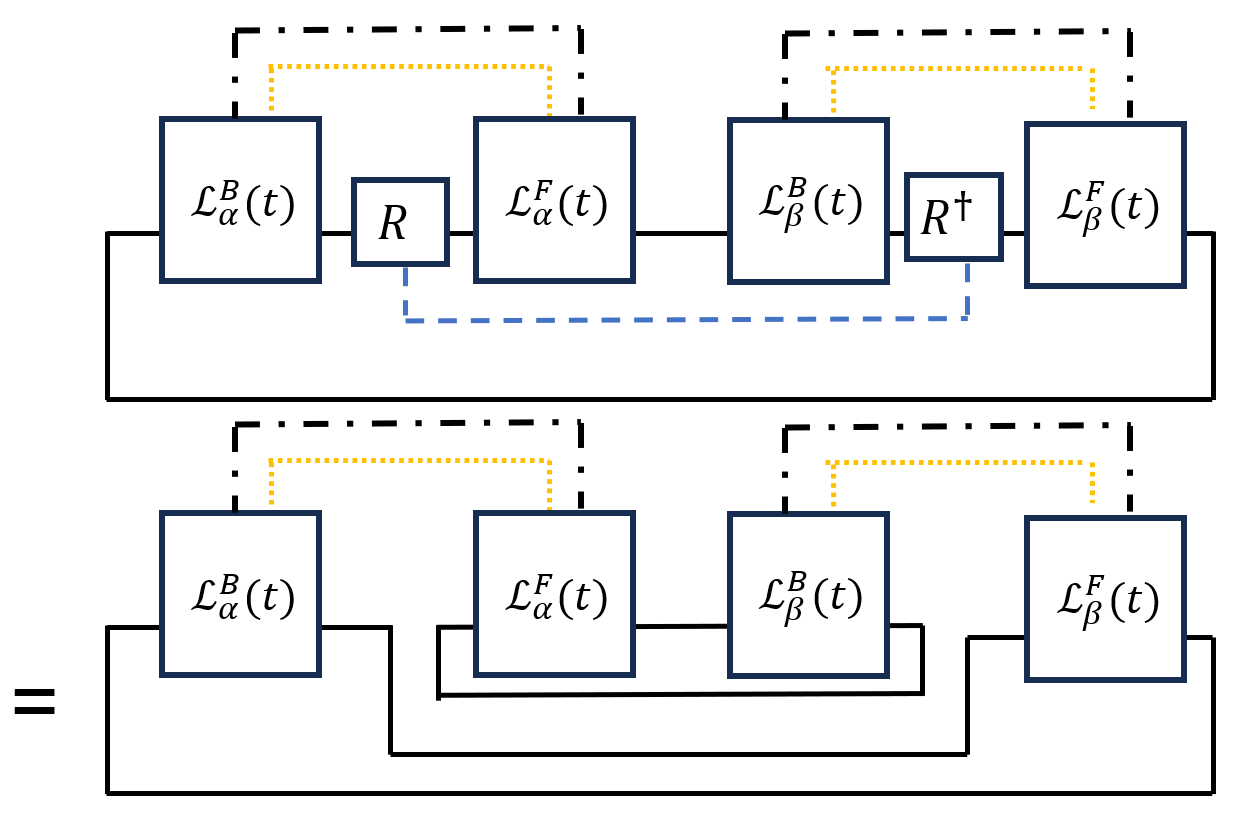} 
		\caption{The diagram representation of the average of OTOC over operator $R$. The black dashed-dot line indicates the correlations between the forward(F) and backward(B) blocks of the Lindblad evolution.}
		\label{open_averB}
\end{figure}

Then, we do the average over $R_B$. We have shown it in the  Fig.~\ref{open_averB}.
We have decomposed the Lindblad evolution into two blocks, denoted as $\mathcal{L}^F$ and $\mathcal{L}^B$ (to facilitate comparison with the unitary evolution in closed systems, we use F to denote \textit{forward} and B to denote \textit{backward}). The black dashed-dot line indicates that there are correlations between the F and B blocks in the Lindblad evolution.

Examining the last line of Fig.~\ref{open_averB}, we find that it takes the form of the unnormalized LE in the open system.
\begin{equation}
		\overline{F^D(t)} 
        =\frac{1}{d_B^2}\overline{\Tr_B \left[ e^{i(H^{D \dagger}_B+V_{\alpha}^{\dagger})t}  e^{-i(H^D_B+V_{\beta })t}\right]}.
\end{equation} 
where the forward and backward lindblad evolutions are represented as $\mathcal{L}_{\alpha}$ and $\mathcal{L}_{\beta}$. Additionally, when we replace the Lindblad evolution with the unitary evolution of the closed system, we find that this diagram reverts to the form of the LE for the closed system, as depicted in the last line of Fig.~\ref{otoc_AB_average}.

\section{The LE dynamics of the dissipative XXZ model}
In this appendix, we provide the results of the LE dynamics in the dissipative XXZ model. And we will see that the structure of the open system LE is not model-dependent, as we will summarize later. The Hamiltonon of the  1D XXZ model is
\begin{equation}
H = J \sum_j \left( S^x_j S^x_{j+1} + S^y_j S^y_{j+1} + \Delta S^z_j S^z_{j+1} \right).
\end{equation}
This model has been realized in several platforms, for example, in ultracold atoms in 1D optical lattices (with spin encoded in two hyperfine states) and in various solid-state or superconducting-qubit implementations. Unlike the SYK model, which has all-to-all random couplings, the XXZ model has only nearest-neighbor interactions and is therefore more experimentally friendly.

We consider local dephasing as the source of dissipation, with Lindblad jump operators $L_j = S^z_j$ on each site. The Lindbladian evolution is thus
\begin{equation} 
		\frac{\partial{\rho}}{\partial t}=-i[H_{\mathrm{xxz}},\rho] +2\gamma\sum_j S_j^z \rho S^{z\dagger}_j-\gamma\sum_j\lbrace S^{z\dagger}_j S_j^z, \rho\rbrace.
	\end{equation}
The numerical simulation of this model also shows the one-minimum structure for the weak dissipation region, and two-local-minimum structure for the strong dissipation region.

\begin{figure}
    \centering
    \includegraphics[width=0.5\linewidth]{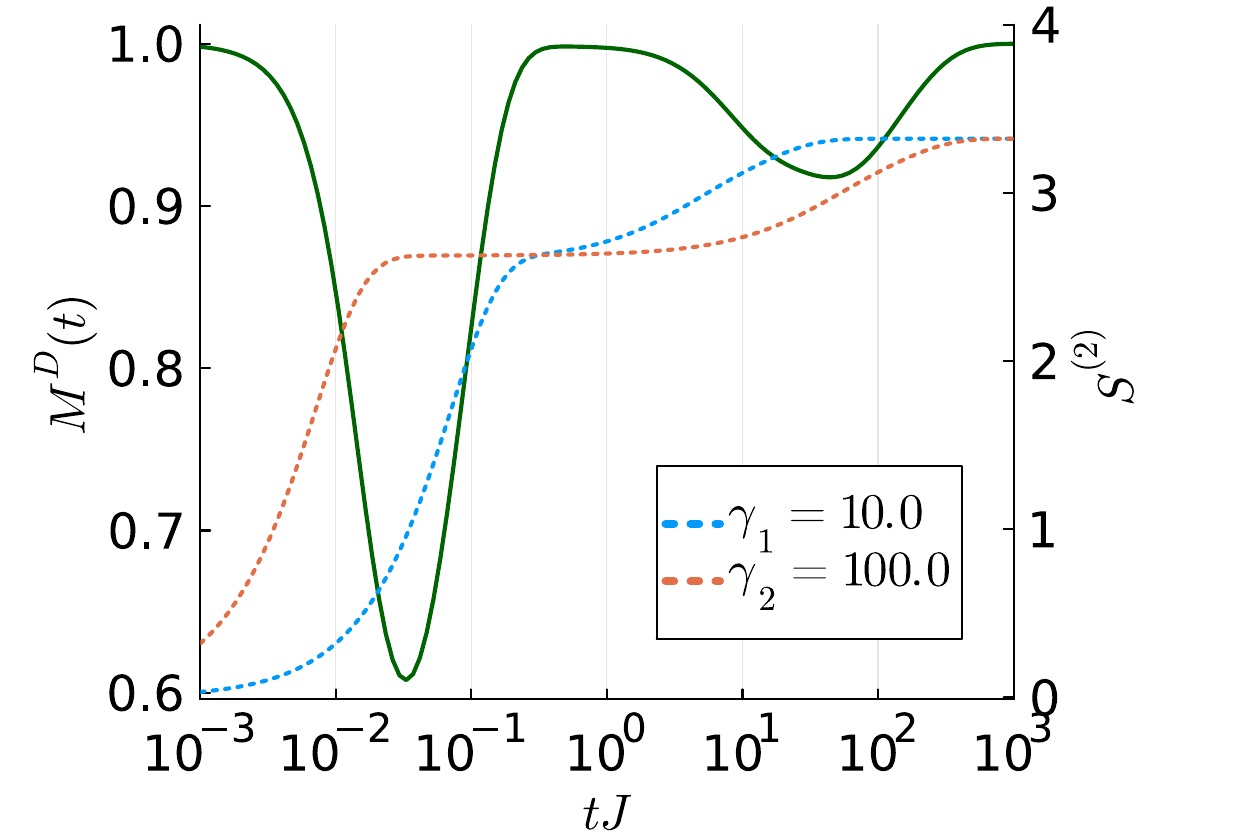}
    \caption{The Loschmidt echo dynamics of the dissipative XXZ model in the strong-dissipation regime. We choose $N = 6$, $q = 3$, $\Delta/J = 2$, $\gamma_1/J = 10$, and $\gamma_2/J = 100$, with the initial state taken to be the ground state of the XXZ Hamiltonian. In this parameter regime, the XXZ model lies in the Néel antiferromagnet phase. Dissipation is applied on all sites via $L_j = S_j^z$, which leads to a degenerate ground space of $H_d$. For comparison, the second R\'enyi entropies for the $\gamma_1$ and $\gamma_2$ evolutions are shown as blue and red dotted lines, respectively. In this regime, the Loschmidt echo exhibits a clear two–local-minima structure.}
    \label{Bfig:XXZ_large_Delta_2.0}
\end{figure}

\begin{figure}
    \centering
    \includegraphics[width=0.5\linewidth]{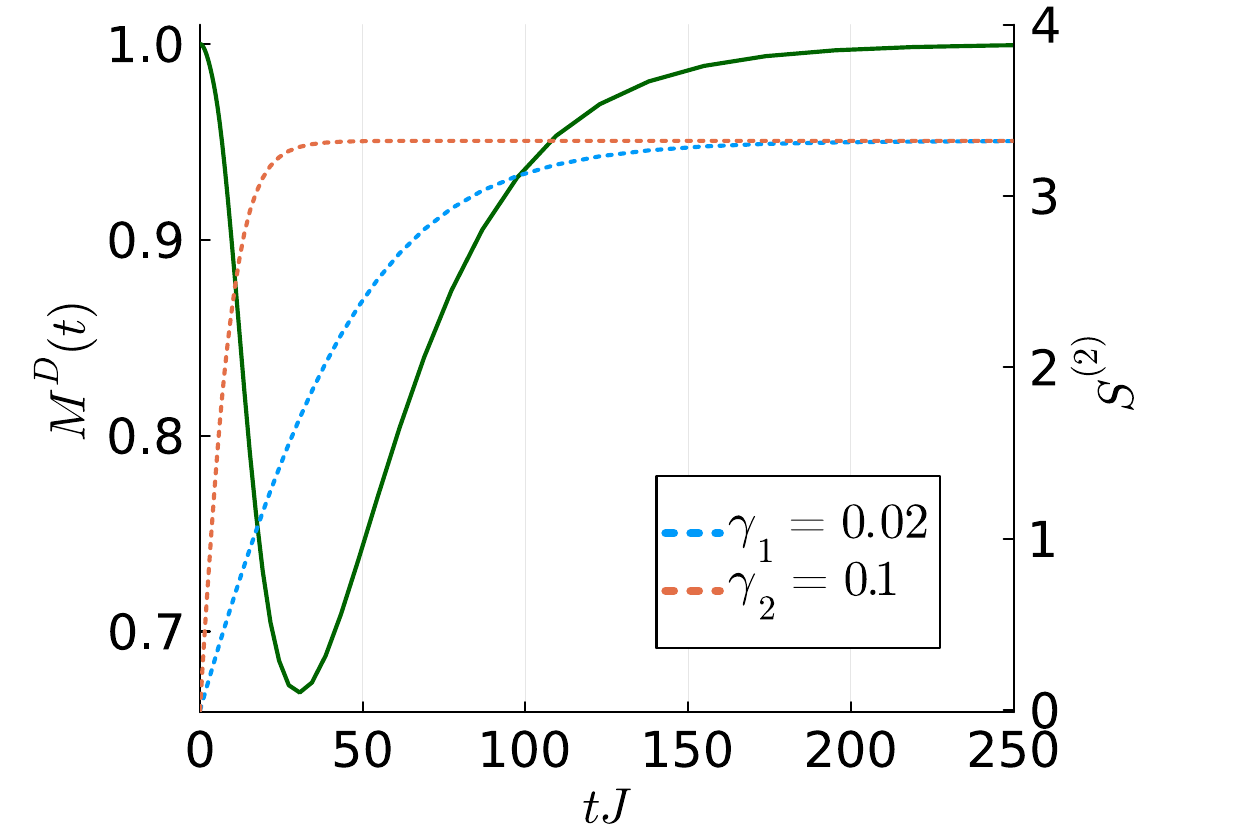}
    \caption{The Loschmidt echo dynamics of the dissipative XXZ model in the weak-dissipation regime. We choose $N = 6$, $q = 3$, $\Delta/J = 2$, $\gamma_1/J = 0.02$, and $\gamma_2/J = 0.1$, with dissipation applied to all sites via $L_j = S_j^z$. In this parameter regime, the XXZ model lies in the N\'eel antiferromagnetic phase. For comparison, the second R\'enyi entropies of the $\gamma_1$ and $\gamma_2$ evolutions are shown as blue and red dotted lines, respectively. The initial state is the ground state of the XXZ Hamiltonian. The minimum of the Loschmidt echo occurs at approximately the same time that the second R\'enyi entropy for the smaller dissipation strength $\gamma_1$ saturates to its late-time plateau.
}
    \label{Bfig:XXZ_small_Delta_2.0}
\end{figure}

\begin{figure}
    \centering
    \includegraphics[width=0.5\linewidth]{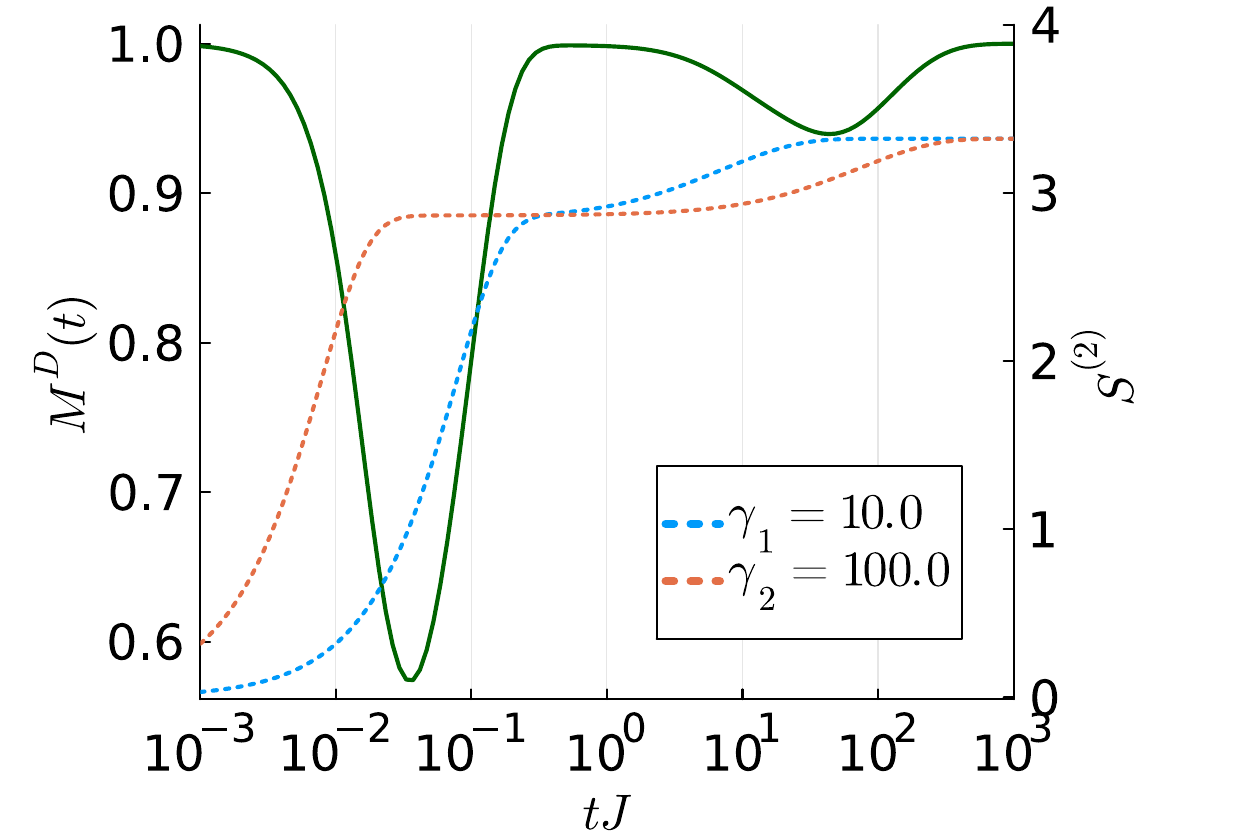}
    \caption{The Loschmidt echo dynamics of the dissipative XXZ model in the strong-dissipation regime. We choose $N = 6$, $q = 3$, $\Delta/J = -1.5$, $\gamma_1/J = 10$, and $\gamma_2/J = 100$, with the initial state taken to be the ground state of the XXZ Hamiltonian. In this parameter regime, the XXZ model lies in the Ferromagnetic phase. Dissipation is applied on all sites via $L_j = S_j^z$, which leads to a degenerate ground space of $H_d$. For comparison, the second R\'enyi entropies for the $\gamma_1$ and $\gamma_2$ evolutions are shown as blue and red dotted lines, respectively. In this regime, the Loschmidt echo exhibits a clear two–local-minima structure.}
    \label{Bfig:XXZ_large_Delta_-1.5}
\end{figure}

\begin{figure}
    \centering
    \includegraphics[width=0.5\linewidth]{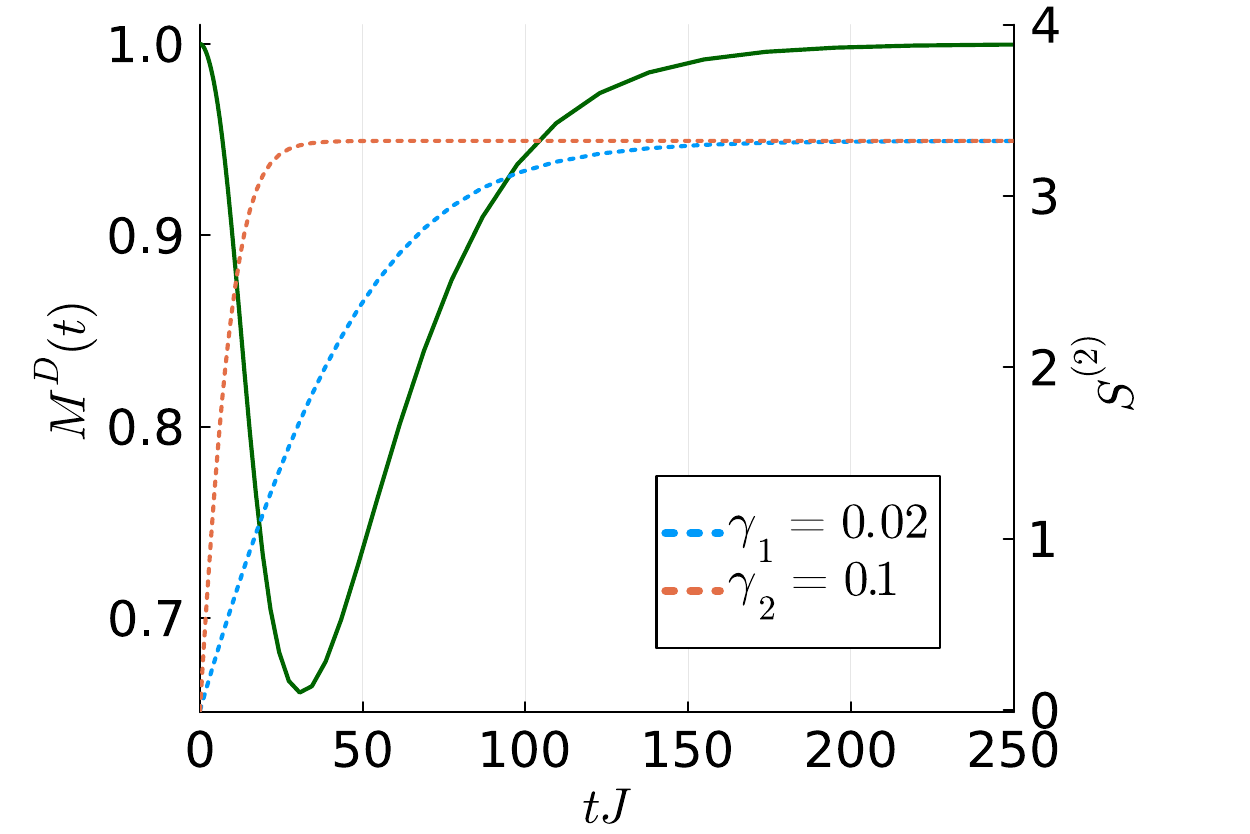}
    \caption{The Loschmidt echo dynamics of the dissipative XXZ model in the weak-dissipation regime. We choose $N = 6$, $q = 3$, $\Delta/J = -1.5$, $\gamma_1/J = 0.02$, and $\gamma_2/J = 0.1$, with dissipation applied to all sites via $L_j = S_j^z$. In this parameter regime, the XXZ model lies in the Ferromagnetic phase. For comparison, the second R\'enyi entropies of the $\gamma_1$ and $\gamma_2$ evolutions are shown as blue and red dotted lines, respectively. The initial state is the ground state of the XXZ Hamiltonian. The minimum of the Loschmidt echo occurs at approximately the same time that the second R\'enyi entropy for the smaller dissipation strength $\gamma_1$ saturates to its late-time plateau.}
    \label{Bfig:XXZ_small_Delta_-1.5}
\end{figure}

\begin{figure}
    \centering
    \includegraphics[width=0.5\linewidth]{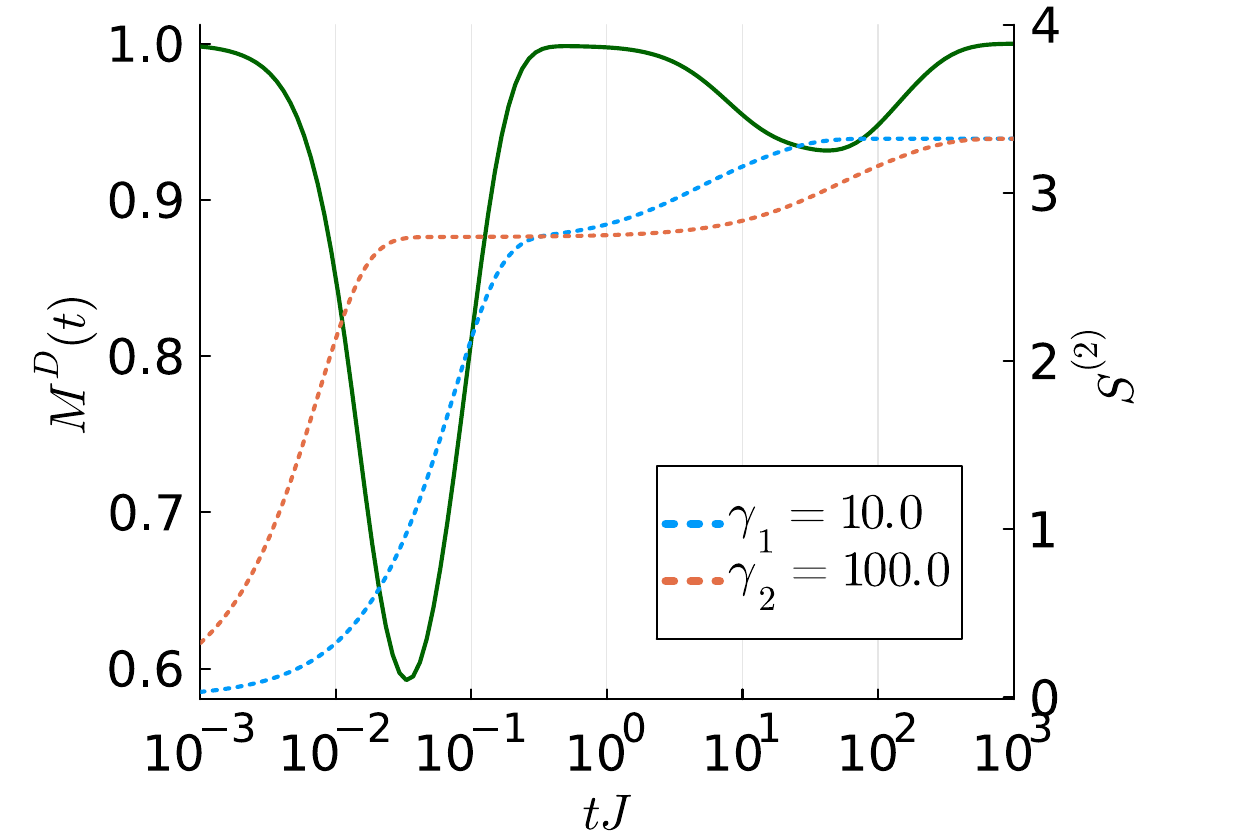}
    \caption{The Loschmidt echo dynamics of the dissipative XXZ model in the strong-dissipation regime. We choose $N = 6$, $q = 3$, $\Delta/J = 0.5$, $\gamma_1/J = 10$, and $\gamma_2/J = 100$, with the initial state taken to be the ground state of the XXZ Hamiltonian. In this parameter regime the XXZ model lies in the gapless Luttinger-liquid phase. Dissipation is applied on all sites via $L_j = S_j^z$, which leads to a degenerate ground space of $H_d$. For comparison, the second R\'enyi entropies for the $\gamma_1$ and $\gamma_2$ evolutions are shown as blue and red dotted lines, respectively. In this regime, the Loschmidt echo exhibits a clear two–local-minima structure.
}
    \label{Bfig:XXZ_large_Delta_0.5}
\end{figure}

\begin{figure}
    \centering
    \includegraphics[width=0.5\linewidth]{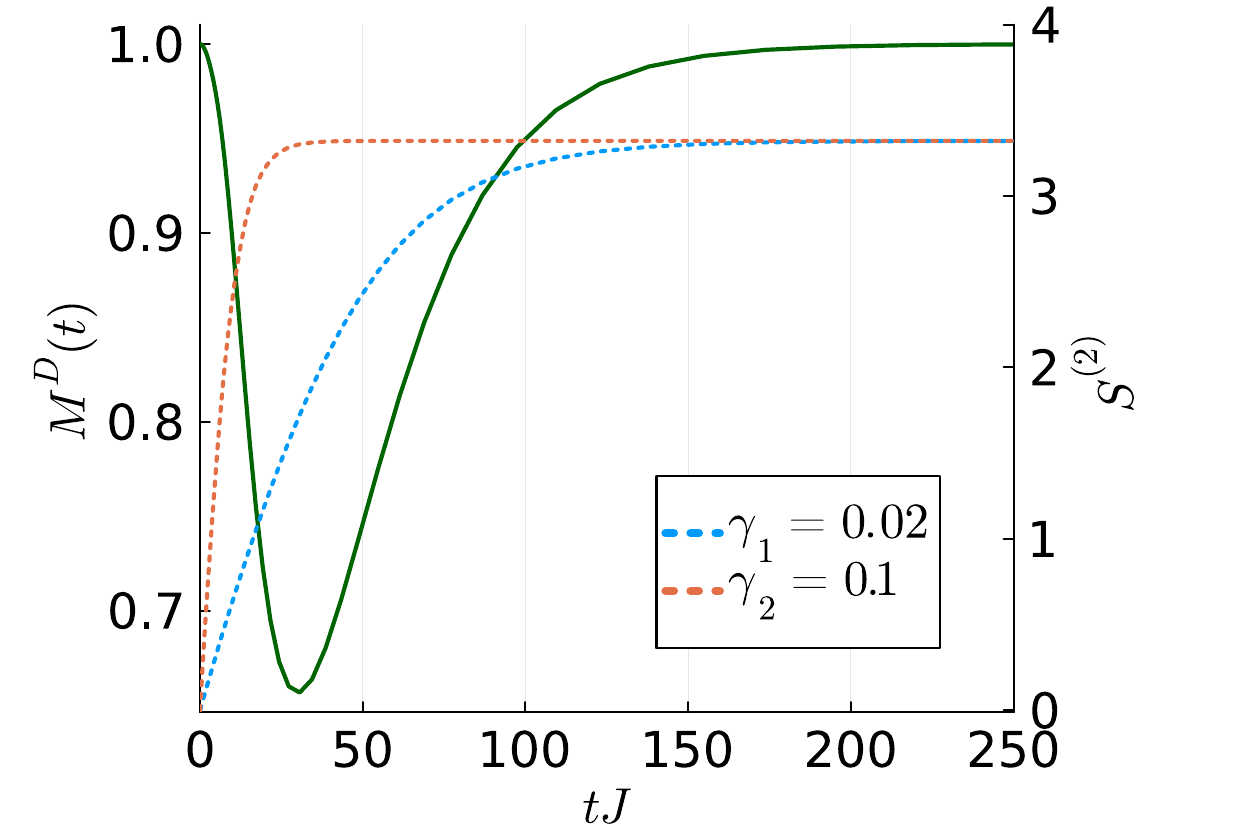}
    \caption{The Loschmidt echo dynamics of the dissipative XXZ model in the weak-dissipation regime. We choose $N = 6$, $q = 3$, $\Delta/J = 0.5$, $\gamma_1/J = 0.02$, and $\gamma_2/J = 0.1$, with dissipation applied to all sites via $L_j = S_j^z$. In this parameter regime, the XXZ model lies in the gapless Luttinger liquid phase. For comparison, the second R\'enyi entropies of the $\gamma_1$ and $\gamma_2$ evolutions are shown as blue and red dotted lines, respectively. The initial state is the ground state of the XXZ Hamiltonian. The minimum of the Loschmidt echo occurs at approximately the same time that the second R\'enyi entropy for the smaller dissipation strength $\gamma_1$ saturates to its late-time plateau.}
    \label{Bfig:XXZ_small_Delta_0.5}
\end{figure}

More concretely, for the parameters shown in Fig.~\ref{Bfig:XXZ_small_Delta_2.0} (weak dissipation, $\Delta/J = 2$, $\gamma_1/J = 0.02$, and $\gamma_2/J = 0.1$), the open-system LE exhibits a single minimum, in direct analogy with the SYK results. In the strong-dissipation regime, Fig.~\ref{Bfig:XXZ_large_Delta_2.0} ($\Delta/J = 2$, $\gamma_1/J = 10$, and $\gamma_2/J = 100$) shows a clear two–local-minima structure, again analogous to the SYK case when the ground space of $H_d$ is degenerate. We further explore different phases of the XXZ model: the gapped ferromagnetic phase for $\Delta/J = -1.5$ [Figs.~\ref{Bfig:XXZ_large_Delta_-1.5} and \ref{Bfig:XXZ_small_Delta_-1.5}] and the gapless Luttinger-liquid phase for $\Delta/J = 0.5$ [Figs.~\ref{Bfig:XXZ_large_Delta_0.5} and \ref{Bfig:XXZ_small_Delta_0.5}]. In all these cases, the LE dynamics in the weak- and strong-dissipation regimes displays the same one-minimum and two–local-minima structures, respectively.

The LE behavior we discuss in our paper is therefore not specific to the SYK model; it is quite general. In the weak-dissipation regime, the single-minimum structure of the LE dynamics follows directly from a perturbative analysis and does not rely on any special property of the SYK Hamiltonian.

In the strong-dissipation regime, the two-local-minima structure of the LE originates from the segmented structure of the Lindblad spectrum. In this regime, the Hamiltonian can be treated as a perturbation. When it does not commute with the dissipative part, this perturbation lifts the degeneracy of the eigenstates of the dissipative generator, leading to a segmented Lindblad spectrum. As a result, the short-time dynamics is controlled by the $\gamma t$ scale, while the long-time dynamics is controlled by the $J^2 t / \gamma$ scale, which gives rise to the two-local-minima structure of the open-system LE.

In short, the emergence of two local minima in the open-system LE only requires that (i) the Hermitian part of the doubled-space Hamiltonian does not commute with the dissipative part, and (ii) the ground space of the dissipative part of the doubled-space Hamiltonian is degenerate (for example, the general local and translation invariant Lindblad operator satisfies this). These conditions are not special to SYK models, but are universal for generic quantum systems under dissipation.

Our numerical simulations for this model show the same qualitative LE behavior as in the SYK case: a single minimum in the weak-dissipation regime and a two–local-minima structure in the strong-dissipation regime.

Overall, the LE behavior we discuss is therefore not specific to SYK, but reflects general dynamical features of open quantum systems with a common infinite-temperature steady state. In particular, in the weak-dissipation regime the single minimum follows from a perturbative analysis and does not rely on any special property of the SYK Hamiltonian, while in the strong-dissipation regime the appearance of two local minima can be traced back to the spectral structure of the doubled-space generator under conditions that are again not unique to SYK.

\end{appendix}

\bibliography{ref.bib}

\end{document}